\documentclass[aps,nofootinbib,preprintnumbers,superscriptaddress]{revtex4}
\usepackage{epsfig}
\usepackage{amssymb,amsmath,bm,bbm}
\usepackage{graphicx}
\usepackage{verbatim}
\usepackage{appendix}
\usepackage[usenames,dvipsnames]{xcolor}

\def\dd{\mathrm{d}}
\def\Mpl{M_{\rm Pl}}

\begin{document}

\preprint{YITP-20-130, IPMU20-0108}

\title{Partial UV Completion of $P(X)$ from a Curved Field Space}

\author{Shinji Mukohyama}
\email{shinji.mukohyama@yukawa.kyoto-u.ac.jp}
\affiliation{Center for Gravitational Physics, Yukawa Institute for Theoretical Physics, Kyoto University, 606-8502, Kyoto, Japan}
\affiliation{Kavli Institute for the Physics and Mathematics of the Universe (WPI), The University of Tokyo, Kashiwa, Chiba 277-8583, Japan}

\author{Ryo Namba}
\email{ryo_namba@sjtu.edu.cn}
\affiliation{Tsung-Dao Lee Institute, Shanghai Jiao Tong University, Shanghai 200240, China}

\begin{abstract}
The $k$-essence theory is a prototypical class of scalar-field models that already gives rich phenomenology and has been a target of extensive studies in cosmology. General forms of shift-symmetric $k$-essence are known to suffer from formation of caustics in a planar-symmetric configuration, with the only exceptions of canonical and DBI-/cuscuton-type kinetic terms. With this in mind, we seek for multi-field caustic-free completions of a general class of shift-symmetric $k$-essence models in this paper. The field space in UV theories is naturally curved, and we introduce the scale of the curvature as the parameter that controls the mass of the heavy field(s) that would be integrated out in the process of EFT reduction.
By numerical methods, we demonstrate that the introduction of a heavy field indeed resolves the caustic problem by invoking its motion near the would-be caustic formation.
We further study the cosmological application of the model. By expanding the equations with respect to the curvature scale of the field space, we prove that the EFT reduction is successfully done by taking the limit of infinite curvature, both for the background and perturbation, with gravity included. The next leading-order computation is consistently conducted and shows that the EFT reduction breaks down in the limit of vanishing sound speed of the perturbation.
\end{abstract}

\maketitle

\section{Introduction}

Scalar fields play important roles in modern cosmology, in both early and late epochs. In the inflationary scenario of the early universe the graceful exit from a quasi-de Sitter expansion requires breaking of the temporal diffeomorphism invariance and correspondingly the introduction of an inflaton, i.e.~a field recording the time remaining before the end of the quasi-de Sitter expansion. Usually, the inflaton is chosen to be a scalar field or a combination of scalar fields. On the other hand, while the late-time acceleration does not necessarily require the same type of symmetry breaking pattern, it is usually considered necessary to introduce extra degrees of freedom if one seeks the origin of the acceleration other than the cosmological constant/vacuum energy. In this case the simplest choice is to introduce an extra degree of freedom via a scalar field. In either epoch, from the viewpoint of the effective field theory (EFT), if a part of the diffeomorphism invariance is broken at the cosmological scale, then there is a priori no reason why the speed limit of a scalar field that plays cosmological roles should agree with the speed of light. For this reason, the kinetic term of a scalar field considered in modern cosmology is often non-canonical. The leading operators in the context of EFT of single-field inflation/dark energy~\cite{Creminelli:2006xe,Cheung:2007st,Creminelli:2008wc} are captured by the action of the form $\int \dd^4x\sqrt{-g}P(\varphi, X)$, provided that the background is sufficiently away from $P_{X}(\varphi,X)=0$,%
\footnote{On the other hand, if the system enjoys the shift symmetry (or an approximate shift symmetry) and if $P_X$ admits a positive root, then $P_X=0$ (or $P_X\approx 0$) is an attractor. When the background is sufficiently close to the attractor $P_X=0$, the sound speed becomes so small that a higher dimensional operator dominates what is usually the dominant gradient term and thus the fluctuations are described by the EFT of ghost condensate~\cite{ArkaniHamed:2003uy,ArkaniHamed:2003uz} or the scordatura theory~\cite{Motohashi:2019ymr,Gorji:2020bfl}.}
where $\varphi$ is a scalar field, $X=-g^{\mu\nu}\partial_{\mu}\varphi\partial_{\nu}\varphi/2$ and a subscript $X$ denotes derivative with respect to $X$. In this sense the $P(\varphi,X)$ model, often called a $k$-essence~\cite{ArmendarizPicon:1999rj,ArmendarizPicon:2000ah,Chiba:1999ka,ArmendarizPicon:2000dh}, is a prototype of a scalar field theory in the context of cosmology.

It is known that the $P(\varphi,X)$ model invariant under an arbitrary constant shift of $\varphi$, i.e.~the $P(X)$ model often called a shift-symmetric $k$-essence, is equivalent to a vorticity-free perfect fluid with the following parametric form of a barotropic equation of state,
\begin{equation}
 \rho = 2P_X(X)X-P(X)\,, \qquad p = P(X)\,, \label{eqn:rho-P-kessence}
\end{equation}
where $\rho$ and $p$ are the energy density and the pressure. In the context of fluid dynamics it has been known that a fluid with a generic equation of state tends to form caustics. By employing techniques developed in the research of partial differential equations and fluid dynamics~\cite{Courant:1948,Lax:1954gzy,Lax:1957hec,Jeffrey:1964yda}, it was shown in \cite{Babichev:2016hys} that the $P(X)$ model generically forms caustics, where the second and higher derivatives of the scalar field diverge. Initially in \cite{Babichev:2016hys}, only the canonical scalar field with $P(X)=AX$, where $A$ is constant, was identified as a model in which the so called simple waves do not lead to caustics. Later in \cite{Mukohyama:2016ipl}, it was found that the Dirac-Born-Infeld (DBI) model with $P(X)=\sqrt{AX+B}$, where $A$ and $B$ are constants, is also free from caustics as far as simple waves are concerned (see also \cite{deRham:2016ged,Tanahashi:2017kgn,Pasmatsiou:2017vcw}).%
\footnote{If the shift symmetry on $\varphi$ is abandoned and $P(X)$ is multiplied by a function of $\varphi$, the resultant non-shift-symmetric DBI would in general form caustics \cite{Felder:2002sv}, which can be interpreted in the stringy setups as the relative difference in the light cone structure between the effective metric for open strings and that for closed ones \cite{Mukohyama:2002vq}.}
This latter case in fact contains the so-called cuscuton model \cite{Afshordi:2006ad,Afshordi:2007yx,Afshordi:2009tt} as one of the limits, $B \to 0$ \cite{deRham:2016ged}.

For a generic $P(X)$ model in which simple waves form caustics, before the formation of caustics the system exits the regime of validity of the EFT and should be taken over by a more fundamental description, i.e. a (partial) UV completion. For example, as suggested in \cite{Babichev:2016hys} and elaborated in \cite{Babichev:2017lrx,Babichev:2018twg}, the $P(X)$ model may emerge as a low-energy effective description of a two-field model with canonical kinetic terms when one of the fields is integrated out. Obviously, in this situation the $P(X)$ description is valid only when second and higher derivatives of the field are sufficiently small in the unit of the mass of the extra field that is integrated out. However, for the two-field models studied in \cite{Babichev:2017lrx,Babichev:2018twg}, once the field space metric is required to be regular at the origin (so that there is no conical singularity) and the range of $\varphi$ is kept non-vanishing (e.g.~$2\pi$), the mass of the extra field is determined by the form of $P(X)$ itself. In a way the mass is solely controlled by the dynamics of the low-energy physics, and it is therefore somewhat contrived to manage the regime of validity of the $P(X)$ description for general $P(X)$.

One of the purposes of the present paper is to extend the two-field model of \cite{Babichev:2017lrx,Babichev:2018twg} so that the mass of the extra field can be made arbitrarily heavy for a given form of $P(X)$. This goal is achieved by promoting the two-dimensional field space, which was taken to be flat in the previous works, to a curved one and by considering the curvature scale of the field space as a parameter that controls the mass of the extra field. In particular, the hyperbolic field space is maximally symmetric and inferred by the so called distance conjecture~\cite{Ooguri:2006in}, which is one of the most conservative cases among all swampland conjectures proposed so far, and thus may be ideal as an ingredient of a possible (partial) UV completion of the $P(X)$ model. We shall also discuss relations to the two-field models studied in the literature~\cite{Tolley:2009fg,Elder:2014fea,Mizuno:2019pcm,Solomon:2020viz}.
As noted in \cite{Mukohyama:2016ipl}, adding higher-order Horndeski terms to $P(X)$ does not ameliorate the caustic problem, and thus we here focus on resolving the issue in the models described by the Lagrangian scalar $P(X)$ without the higher-order terms.

Another purpose of the present paper is to see how the extra field behaves as the system approaches the incident of a caustic formation. For this purpose we study a planar symmetric configuration of the two-field model in the Minkowski spacetime. We employ numerical methods to integrate the full two-field system of nonlinear coupled equations as well as the corresponding equations in a single-field $P(X)$ model, which is essentially obtained by the Legendre transformation from the two-field model with one field being infinitely massive. The observation of caustics in the original single-field model and of its resolution in the corresponding two-field completion evidently shows the validity of our approach, which is the first explicit numerical demonstration to our knowledge. The result also gives a clear illustration of the necessity of the second field in order to avoid the caustic formation.

In view of further applications in realistic setups, we also study cosmology in the obtained two-field model minimally coupled to General Relativity. We expand the equations of motion by the curvature scale of the field space so that the deviation from the $P(X)$ description can be systematically studied. It is then shown that both at the level of the Friedmann–Lema\^{i}tre–Robertson–Walker (FLRW) background and at the level of linear perturbations, the $P(X)$ description is valid at energies and momenta sufficiently lower than the mass of the extra field that is controlled by the curvature of the field space.
We then proceed to the next order in the expansion and illustrate how the higher-order equations can be systematically obtained by iteratively integrating out the heavy field. It is found that the cutoff scale of the single-field EFT is related to the sound speed of the perturbation and, in particular, the EFT expansion with respect to the curvature scale would break down in the limit of vanishing sound speed, the finding consistent with a generic expectation \cite{Cheung:2007st}.

The rest of the paper is organized as follows.
In Sec.~\ref{sec:twofield}, we introduce the class of $P(X)$ models that we study and then promote it to those of two fields in a curved field space. We consider both linear-kinetic and DBI-type completions.
We then conduct numerical computation of the obtained model on a planar-symmetric configuration in Sec.~\ref{sec:planar}. This provides an explicit demonstration of the avoidance of caustic formations in the two-field model, observing that the would-be divergence of second derivatives of the light field is smoothed out by the onset of the motion of the heavy field.
In Sec.~\ref{sec:cosmology}, we consider the cosmology of the two-field models, both the FLRW background and the perturbations around it, showing that a consistent expansion in terms of the curvature scale of the field space can be done.
Sec.~\ref{sec:discussion} summarizes our results and discusses their implications.
In Appendix \ref{app:action} we collect some technicalities in changing variables with derivatives involved and obtaining the action of the new variable.

\section{Two-field model}
\label{sec:twofield}

It was demonstrated in \cite{Babichev:2016hys,Mukohyama:2016ipl,deRham:2016ged} that models of the (shift-symmetric) $k$-essence and Horndeski types are in general vulnerable to formations of caustic singularities in the flat spacetime. In \cite{Mukohyama:2016ipl}, it was shown that the classes of models immune to such pathological behaviors consist not only of canonical scalar fields, which had already been shown in \cite{Babichev:2016hys}, but also of the Dirac-Born-Infeld (DBI) scalars, and that this exhausts the list.%
\footnote{In fact, the result in \cite{Mukohyama:2016ipl} already included the so-called cuscuton model ${\cal L} \propto \sqrt{X}$, where $X \equiv - (\partial\varphi)^2/2$, as one of the limits, whose caustic-free nature was explicitly stated in \cite{deRham:2016ged}. The latter reference \cite{deRham:2016ged} extended the analysis to an $SO(p)$ symmetry in an arbitrary number of space dimensions. However, the most restrictive, that is the most conservative, caustic-free condition is found to emerge from the one with a planar-symmetric configuration, as studied in \cite{Mukohyama:2016ipl}.}
Caustic formations do not necessarily imply a breakdown of the evolution of a considered system but rather hint a departure from the validity regime of the $k$-essence/Horndeski model used as an effective theory. Before caustics form, operators in a more fundamental theory that are integrated out in the effective description start being in action.
In this section, we provide a class of caustic-free completion of $k$-essence models by introducing an additional scalar field.

We aim to complete shift-symmetric $k$-essence by a two-field system with a curved field space that has a line element,
\begin{equation}
\gamma_{IJ} \, \dd \Phi^I \dd \Phi^J = \dd \chi^2 + f(\beta \chi) \, \dd \varphi^2
\; ,
\label{fieldspace}
\end{equation}
where a non-negative function $f$ determines the shape of the field space spanned by $\Phi^I = ( \chi, \varphi )$, and for a fixed function $f$ a constant $\beta$ controls the curvature of the field space, the mass of the extra field and thus the cutoff scale of the corresponding single-field effective field theory (EFT). The curvature associated with the field-space metric $\gamma_{IJ} = {\rm diag} ( 1 , f )$ is quantified by the Ricci tensor, given by ${\cal R}_{IJ} = {\rm diag} \left( - \beta^2 f^{-1/2} (f^{1/2})'' , \, - \beta^2 f^{1/2}(f^{1/2})'' \right)$, where prime denotes derivative with respect to the argument $\beta\chi$. Some classifications of the field space are
\begin{equation}
    \sqrt{f(\beta\chi)} =
    \begin{cases}
    \beta \chi \; , \qquad & \mbox{flat}\  (\mbox{with}\  \beta=1)\; , \vspace{1mm}\\
    \exp( \beta\chi ) \; , \qquad & \mbox{hyperboloidal} \; , \vspace{1mm} \\
    \displaystyle
    \frac{\sin(\beta\chi)}{\beta} \; , \qquad & \mbox{spheroidal} \; ,
    \end{cases}
\end{equation}
where for the flat case $\beta=1$ is required by the avoidance of a conical singularity at $\chi=0$ while for the spheroidal case $\beta^{-1}$ is introduced for the same reason. Our primary interest is the case of nontrivial field space geometry ${\cal R}_{IJ} \ne 0 $, that is $(f^{1/2})'' \ne 0$.
In the following subsections, we therefore consider a linear and DBI-type kinetic terms of two fields whose space geometry is curved according to \eqref{fieldspace}.

\subsection{Two-field model with linear kinetic terms}
\label{subsec:linear}

In this subsection we develop a two-field system with a linear kinetic terms that serves as a (partial) UV completion of the general $P(\varphi,X)$ models. We first consider a simpler case with shift symmetry, i.e.~$P(X)$ models, and then extend it to more general $P(\varphi,X)$ models.

\subsubsection{Equivalent description of $P(X)$}

As a preparation for the construction of a two-field completion of $P(X)$ models, we first rewrite the Lagrangian scalar $P(X)$ as
\begin{equation}
    {\cal L}_{\rm lin\mbox{-}EFT} = f ( \beta \chi ) \, X - V ( \beta \chi ) \; ,
    \label{LagEFT}
\end{equation}
where $V$ is a function to be determined so that ${\cal L}_{\rm lin\mbox{-}EFT}$ reduces to $P(X)$ after integrating out the auxiliary field $\chi$. The value of $\chi$ is determined by its equation of motion that is obtained by taking variation of the action $\int \dd^4x\sqrt{-g}{\cal L}_{\rm lin\mbox{-}EFT}$ with respect to $\chi$,
\begin{equation}
    \frac{\dd v}{\dd f} = X \; .
    \label{chiX_EFT}
\end{equation}
Here, we have considered $V(\beta\chi)$ as a function of $f$ and denoted it as $v(f)$, assuming that $f'( \beta\chi ) \ne 0$ in the range of $\beta\chi$ that is of our interest. By algebraically solving \eqref{chiX_EFT} with respect to $f$, we obtain $f=f(X)$. Plugging this solution into \eqref{LagEFT}, we can rewrite ${\cal L}_{\rm lin\mbox{-}EFT}$ as a function of $X$ and then demand that this function of $X$ coincides with $P(X)$, that is,
\begin{equation}
 f X - v(f) = P(X) \; .
    \label{LagPX}
\end{equation}
In fact, the relation \eqref{LagPX} is interpreted as $v(f)$ being the Legendre transformation of $P(X)$, or $P(X)$ being the Legendre transformation of $v(f)$. This fact immediately results in the following relations: $\dd v/\dd f=X$, which is \eqref{chiX_EFT}, $P_X = f$, and $\dd^2v/\dd f^2=1/P_{XX}$. The invertiblity of the Legendre transformation requires that $P(X)$ be a convex (or concave) function, i.e. $P_{XX}>0$ (or $P_{XX}<0$). It then follows that $v(f)$ is also a convex (or concave) function, i.e. $\dd^2v/\dd f^2>0$ (or $\dd^2v/\dd f^2<0$). In other words, the class of the models given by \eqref{LagEFT} can in principle cover all the shift-symmetric $k$-essence models $P(X)$ that respect $P_{XX}\ne 0$.

By varying $\int \dd^4x\sqrt{-g}{\cal L}_{\rm lin\mbox{-}EFT}$, where ${\cal L}_{\rm lin\mbox{-}EFT}$ is given by \eqref{LagEFT}, with respect to $\varphi$, we find the equation of motion for $\varphi$ as
\begin{equation}
    \nabla^\mu \left( f \nabla_\mu \varphi \right) = 0 \; ,
    \label{EOM_EFT}
\end{equation}
where $\nabla_\mu $ is the covariant derivative associated with the spacetime metric.
By taking derivative of $P(X)$ in \eqref{LagPX} with respect to $X$ with the use of \eqref{chiX_EFT}, we find the correspondence
\begin{equation}
    P_X = f \; ,
    \label{PX_f}
\end{equation}
which was already inferred from the fact that $v(f)$ is a Legendre transformation of $P(X)$, and thus one can easily identify \eqref{EOM_EFT} with the equation of motion for the $k$-essence models. The energy-momentum tensor corresponding to \eqref{LagEFT} is found as
\begin{equation}
    T^{\rm lin\mbox{-}EFT}_{\mu\nu} = f \, \partial_\mu \varphi \, \partial_\nu \varphi + \left( f X - V \right) g_{\mu\nu} \; ,
    \label{EMtensorEFT}
\end{equation}
and the corresponding energy density and pressure can be found by $\rho_{\rm lin\mbox{-}EFT}=n^\mu n^\nu T^{\rm lin\mbox{-}EFT}_{\mu\nu}$ and $P_{\rm lin\mbox{-}EFT}=h^{\mu\nu} T^{\rm lin\mbox{-}EFT}_{\mu\nu} /3 $, respectively, where $n_\mu = \partial_{\mu}\varphi/\sqrt{2X}$ is the unit vector normal to the constant-$\varphi$ hypersurface with the inverse induced metric $h^{\mu\nu}=g^{\mu\nu}+n^{\mu}n^{\nu}$. We find
\begin{equation}
 \rho_{\rm lin\mbox{-}EFT} = 2fX - (fX - V)\,, \quad P_{\rm lin\mbox{-}EFT} = f X - V\,.
\end{equation}
which reproduce \eqref{eqn:rho-P-kessence} upon using \eqref{LagPX} and \eqref{PX_f}.

In this equivalent description of the single-field $P(X)$ model, the parameter $\beta$ has no physical meaning since it can be absorbed by rescaling $\chi$ in \eqref{LagEFT} or \eqref{LagPX} . However, in completing the single-field theory to a two-field UV theory below, it plays a fundamental role by controlling the energy scale $\propto \beta^{-1} $ for which the dynamics of $\chi$ becomes relevant to resolve the caustic singularities, as demonstrated below and in Sec.~\ref{sec:planar}.

\subsubsection{Two-field completion by adding kinetic term for extra scalar}
\label{subsec:canonical}

For a multi-field completion of the effective theory \eqref{LagEFT} by the curved field space \eqref{fieldspace}, we assume that the UV theory above the cutoff scale of the single-field EFT consists of multi scalar fields in a curved field space. Hence our complete Lagrangian scalar takes the form
\begin{equation}
{\cal L}_{\rm lin} = - \frac{1}{2} \, \gamma_{IJ} \nabla_\mu \Phi^I \nabla^\mu \Phi^J - V(\Phi^I)
\; ,
\label{LagMulti}
\end{equation}
where the curvature of $\gamma_{IJ}$ is negative.
To capture the mechanism of our interest, it is sufficient to identify a flat direction in the field space with $\varphi$ and to denote by $\chi$ a representative massive direction. The Lagrangian scalar \eqref{LagMulti} then reduces to a $2$-field one with $\gamma_{IJ} $ identified with the one in \eqref{fieldspace},
\begin{equation}
{\cal L}_{\rm lin} = - \frac{1}{2} \left( \partial \chi \right)^2 - \frac{f(\beta \chi)}{2} \left( \partial \varphi \right)^2 - V(\beta \chi) \; ,
\label{LagTwo}
\end{equation}
where $V$ is independent of a flat direction $\varphi$.%
\footnote{In principle, there can be a mixing kinetic term of the form $\nabla_\mu \varphi \nabla^\mu \chi$, but it can be removed by a field redefinition, at the price of additional terms in the potential. Imposing shift symmetry on the resultant $\varphi$, $V$ is left independent of $\varphi$. Then the kinetic term of $\chi$ can be canonically normalized without loss of generality.}
Now it is evident that $\beta$ is a parameter of mass dimension $-1$ that controls the ``mass'' of the massive mode $\chi$.

The equations of motion for $\chi$ and $\varphi$ are, respectively,
\begin{align}
& - \nabla^2 \chi + \beta \left[ \frac{f'}{2} \left( \partial \varphi \right)^2 + V' \right] = 0 \; ,
\label{EOMTwo_chi}\\
& - \nabla_\mu \left( f \nabla^\mu \varphi \right) = 0 \; .
\label{EOMTwo_phi}
\end{align}
In the limit $\beta \to \infty$, \eqref{EOMTwo_chi} reduces to
\begin{equation}
\frac{V'}{f'} = - \frac{1}{2}\left( \partial \varphi \right)^2 \; ,
\label{ConstraintTwo}
\end{equation}
which is identical to the constraint \eqref{chiX_EFT}.
Since the equation of motion for $\varphi$ in \eqref{EOMTwo_phi} is exactly the same as the effective model \eqref{LagEFT}, it is clear that the dynamics of this system is identical to that of the effective theory \eqref{LagTwo} as long as $\beta \to \infty$ is a justified limit to give \eqref{ConstraintTwo}.
The energy momentum tensor of the full theory \eqref{LagTwo} is
\begin{equation}
T^{\rm lin}_{\mu\nu}
= \partial_\mu \chi \, \partial_\nu \chi
+ f \, \partial_\mu \varphi \, \partial_\nu \varphi
+ g_{\mu\nu} \left[ - \frac{1}{2} \left( \partial\chi \right)^2 - \frac{f}{2} \left( \partial \varphi \right)^2 - V \right] \; .
\label{EMtensorTwo}
\end{equation}
Notice that, in the limit $\beta \to \infty$, the field $\chi$ is a massive mode and thus $\partial \chi \to 0$, in which case the energy momentum tensor in \eqref{EMtensorTwo} also becomes identical to the EFT one \eqref{EMtensorEFT}.
This proves the equivalence between the effective theory \eqref{LagEFT} and its two-field completion \eqref{LagTwo} in the limit $\beta \to \infty$.

\subsubsection{Reconstruction of $v(f)\equiv V(\beta\chi)$}
\label{subsection:reconstruct-potential}

For a given $f$ and $V$, the corresponding $P(X)$ theory can be obtained by \eqref{LagPX}, together with $\beta\chi$ as a function of $X$ that is the solution of \eqref{chiX_EFT}. In order to express the potential $V$ for a given $P(X)$ and the form of $f$ that is determined from the field space curvature as in \eqref{fieldspace}, let us consider a generic expansion of $P(X)$,
\begin{equation}
    P(X) = \sum_n c_n X^n \; ,
    \label{PX_full}
\end{equation}
where $c_n $ are the coefficients of $X$ polynomials. Using \eqref{LagPX} with the relation \eqref{chiX_EFT}, one obtains an equation
\begin{equation}
    v - f^2 \, \frac{\dd v}{\dd f} + \sum_n c_n \left[ \frac{\dd v}{\dd f} \right]^n = 0 \; ,
    \label{V_eq}
\end{equation}
where $V' / f' = \dd v / \dd f$ has been used with the identification $V(\beta\chi) = v(f)$. By solving the differential equation \eqref{V_eq}, the form of $v$, and consequently $V$, can be found.
Eq.~\eqref{V_eq} is a necessary condition for $v$ to satisfy in order to represent a given $P(X)$ model. This is however not sufficient, and as discussed below \eqref{LagPX}, the Legendre transform can be done if and only if $P_{XX} \ne 0$.
The first nontrivial example, which includes up to $X^2 $ terms in $P(X) = c_0 + c_1 X + c_2 X^2$, i.e.~$c_n =0 $ for all $n$ except for $c_0, c_1 $ and $c_2$, yields a solution~%
\footnote{Eq.~\eqref{V_eq} gives another set of solutions
\begin{equation}
    v = - c_0 + C \left( f - c_1 \right) - C^2 c_2 \; ,
    \label{V_general}
\end{equation}
where $C$ is an integration constant. However this does not meet the condition of the Legendre transformation, since the second derivative vanishes, i.e.~$\dd^2 v / \dd f^2 = 0$. Thus these solutions are not taken as appropriate forms of $v(f)$.}
\begin{equation}
    v = - c_0 + \frac{1}{4 \, c_2} \left( f - c_1 \right)^2 \; .
    \label{V_particular}
\end{equation}
From a theoretical/model-building point of view, if a UV theory that takes the form \eqref{LagMulti}, or \eqref{LagTwo}, induces the dynamics similar to a $P(X)$ theory truncated at $X^2$, then the corresponding potential should be identified with the one of the form \eqref{V_particular}.
In Secs.~\ref{sec:planar} and \ref{sec:cosmology}, we focus on the the two-field theory \eqref{LagTwo} for detailed analyses, and in Sec.~\ref{sec:planar} we explicitly utilize the form \eqref{V_particular} for numerical demonstration of caustic formation and resolution.

\subsubsection{Extension to $P(\varphi,X)$}
\label{subsec:completion-generalP}

The extension of the two-field completion of $P(X)$ models to more general $P(\varphi,X)$ models is straightforward. Assume that $P(\varphi,X)$ is convex (or concave) with respect to $X$, i.e. $P_{XX}>0$ (or $P_{XX}<0$), so that the Legendre transformation of $P(\varphi,X)$ with respect to $X$ exists and is unique. Let $v(\varphi,f)$ be the Legendre transformation of $P(\varphi,X)$ with respect to $X$ (and thus $P(\varphi,X)$ be the Legendre transformation of $v(\varphi,f)$ with respect to $f$). The two-field completion with the field space metric \eqref{fieldspace} is then given by the action $\int d^4x\sqrt{-g}{\cal L}_{\rm lin}$, where
\begin{equation}
{\cal L}_{\rm lin} = - \frac{1}{2} \left( \partial \chi \right)^2 - \frac{f(\beta \chi)}{2} \left( \partial \varphi \right)^2 - V(\varphi,\beta \chi) \; ,\quad V(\varphi,\beta\chi)\equiv v(\varphi,f(\beta\chi))\;.
\end{equation}
The single-field $P(\varphi,X)$ model is recovered in the limit $\beta\to\infty$.

\subsection{Two-field model with DBI-type kinetic terms}
\label{subsec:DBI}

In the previous subsection we have developed a (partial) UV completion of the class of $P(X)$ and $P(\varphi,X)$ models by a two-field system that has a linear kinetic terms with the curved field space \eqref{fieldspace}. In the next sections we shall use it to resolve the problem of caustic singularities and study cosmology. Before that, in this subsection we consider another possible (partial) UV completion of a class of $P(X)$ models by the curved field space \eqref{fieldspace} but with Dirac-Born-Infeld (DBI)-type kinetic terms. Extension to $P(\varphi,X)$ is straightforward and discussed in Sec.~\ref{subsubsec:extension_DBI}. Readers who are interested in the resolution of the caustic singularities and cosmological application of the completion with linear kinetic terms may skip this subsection and directly go to the next sections.

As shown in \cite{Mukohyama:2016ipl}, not only the canonical scalar field but also the DBI model with $P(X)=\sqrt{AX+B}$, where $A$ and $B$ are constants, is also free from caustics as far as simple waves are concerned. Also, string theory allows not only scalar fields with linear kinetic terms but also those with DBI-type kinetic terms that stem from D-branes moving in extra dimensions~\cite{Polchinski:1998rr}. For these reasons, it is reasonable to ask whether we can extend the (partial) UV completion of $P(X)$ models to a two-field system with DBI-type kinetic terms. In this subsection we answer this question positively by explicitly constructing such a two-field system.

\subsubsection{Equivalent description of $P(X)$}

In order to construct a two-field system with DBI-type kinetic terms that can (partially) UV-complete a class of $P(X)$ models, as the first step we now consider an single-field EFT of the form
\begin{equation}
    {\cal L}_{\rm DBI\mbox{-}EFT} = - \sqrt{1 - 2 f(\beta\chi) \, X} - V(\beta\chi) \; ,
    \label{LagEFT_DBI}
\end{equation}
where again $X = - (\partial\varphi)^2/2$, $\chi$ is an auxiliary field, $f$ ($>0$) and $V$ are functions of $\beta\chi$, and $\beta$ is a constant. In this single-field description, the value of $\chi$ is determined by the constraint equation
\begin{equation}
 \frac{\dd v}{\dd f} = \frac{X}{\sqrt{1 - 2 fX}} \; ,
    \label{chiX_DBI}
\end{equation}
where again we have regarded $V(\beta\chi)$ as a function of $f$ and denoted it as $v(f)$, assuming that $f'( \beta\chi ) \ne 0$ in the range of $\beta\chi$ that is of our interest. Solving this for $f$ in terms of $X$ and plugging it back into \eqref{LagEFT_DBI} results in a class of $P(X)$ theories. We determine $v(f)$ such that
\begin{equation}
 -\sqrt{1-2fX} - v(f) = P(X)\,.
 \label{LagPX_DBI}
\end{equation}
As stated below \eqref{LagPX}, the parameter $\beta$ carries no physical meaning at this stage, as it can be absorbed in the redefinition of $\chi$. Its role as the controlling cutoff scale of the EFT will soon be clear once its two-field completion is introduced below.
The equation of motion and the energy-momentum tensor associated with \eqref{LagEFT_DBI} are, respectively,
\begin{align}
    \nabla^\mu \left( \frac{f \nabla_\mu \varphi}{\sqrt{1 - 2 f X}} \right) = 0 \; , \qquad
    T^{\rm DBI\mbox{-}EFT}_{\mu\nu} = \frac{f \, \partial_\mu \varphi \, \partial_\nu \varphi}{\sqrt{1 - 2 f X}}
    + g_{\mu\nu} \left( - \sqrt{1 - 2 fX} - V \right) \; .
    \label{EOM_EMT_DBI}
\end{align}
By observing $P_X = f / \sqrt{1 - 2 fX}$ by the use of \eqref{chiX_DBI}, we identify the system governed by the above E.o.M.~and energy-momentum tensor with those of the corresponding $P(X)$ theory.

\subsubsection{Two-field completion by adding kinetic term for extra scalar}

Our two-field completion of \eqref{LagEFT_DBI} by the curved field space \eqref{fieldspace} is done by the inclusion of the kinetic term of $\chi$ in the following manner:
\begin{align}
    {\cal L}_{\rm DBI} &
    = - \sqrt{1 + \gamma_{IJ} \nabla_\mu \Phi^I \nabla^\mu \Phi^J} - V(\Phi^I)
    \nonumber\\ &
    = - \sqrt{1 + \left( \partial\chi \right)^2 + f(\beta\chi) \left( \partial\varphi \right)^2} - V(\beta\chi) \; ,
    \label{LagTwo_DBI}
\end{align}
keeping the $\varphi$ direction shift-symmetric.
The dynamics of this two-field system is governed by the equations of motion
\begin{align}
    &
    - \nabla_\mu \left[ \frac{\nabla^\mu \chi}{\sqrt{1 + \left( \partial\chi \right)^2 + f \left( \partial\varphi \right)^2}} \right]
    + \beta \left[ \frac{1}{2}\frac{f' \left( \partial\varphi \right)^2}{\sqrt{1 + \left( \partial\chi \right)^2 + f \left( \partial\varphi \right)^2}}
    + V' \right] = 0 \; ,
    \label{EOMTwo_DBI_chi}
    \\ &
    - \nabla_\mu \left[ \frac{f \nabla^\mu \varphi}{\sqrt{1 + \left( \partial\chi \right)^2 + f \left( \partial\varphi \right)^2}} \right] = 0 \; .
    \label{EOMTwo_DBI_phi}
\end{align}
In the limit $\beta \to \infty$, \eqref{EOMTwo_DBI_chi} gives $\chi \propto \beta^{-1}$, and thus \eqref{EOMTwo_DBI_chi} and \eqref{EOMTwo_DBI_phi} exactly reduce to the constraint \eqref{chiX_DBI} and the E.o.M.~\eqref{EOM_EMT_DBI}, respectively, in the single-field EFT.
The energy-momentum tensor associated with the Lagrangian scalar \eqref{LagTwo_DBI} is
\begin{align}
    T^{\rm DBI}_{\mu\nu} = \frac{\partial_\mu \chi \, \partial_\nu \chi + f \, \partial_\mu \varphi \, \partial_\nu \varphi}{\sqrt{1 + \left( \partial\chi \right)^2 + f \left( \partial\varphi \right)^2}}
    + g_{\mu\nu} \left[ - \sqrt{1 + \left( \partial\chi \right)^2 + f \left( \partial\varphi \right)^2} - V \right] \; ,
\end{align}
and reduces to $T^{\rm DBI\mbox{-}EFT}_{\mu\nu}$ in \eqref{EOM_EMT_DBI} in the limit $\beta \to \infty$.
Therefore, the two-field theory \eqref{LagTwo_DBI} derives the single-field one \eqref{LagEFT_DBI} as its single-field EFT limit, even with the inclusion of (minimally coupled) gravity.

\subsubsection{Reconstruction of $v(f)\equiv V(\beta\chi)$}

The form of the potential $V$ for a given $f$, i.e. $v(f)=V(\beta\chi)$, with the corresponding single-field EFT of the form \eqref{PX_full}, can be obtained by solving a differential equation similar to \eqref{V_eq}, that is
\begin{equation}
    v - f \, \frac{\dd v}{\dd f} \pm \sqrt{1 + f \left( \frac{\dd v}{\dd f} \right)^2}
    + \sum_n c_n \left[ - f \left( \frac{\dd v}{\dd f} \right)^2 \pm \sqrt{1 + f^2 \left( \frac{\dd v}{\dd f} \right)^4} \right]^n = 0 \; .
\end{equation}
This apparently does not admit a closed analytical solution for a general $P(X)$, but numerical reconstruction of $V$ is straightforward.

\subsubsection{Extension to $P(\varphi,X)$}
\label{subsubsec:extension_DBI}

We discuss the extension of the two-field completion with DBI-type kinetic terms in order to accommodate $P(\varphi, X)$ models in this subsubsection. While, unlike the case with linear kinetic terms, the relation \eqref{LagPX_DBI} does not host an immediate interpretation as a Legendre transformation due to the square root structure of DBI, the extension itself is straightforward.
We promote the function $V$ in \eqref{LagEFT_DBI} to the one on both $\beta\chi$ and $\varphi$, i.e.~$V(\varphi, \beta\chi)$. Then the constraint equation \eqref{chiX_DBI} is modified to be
\begin{equation}
    \frac{\partial v}{\partial f} = \frac{X}{\sqrt{1 - 2 f X}} \; ,
\end{equation}
where $v(\varphi , f)$ is $V$ regarded as a function of $\varphi$ and $f$, provided $f'(\beta\chi) \ne 0$. This equation can be algebraically solved to give the relation of $f$ to $\varphi$ and $X$. Then we can rewrite the action as a function of $\varphi$ and $X$ and then require that it coincide with $P(\varphi,X)$, i.e.
\begin{equation}
    - \sqrt{1 - 2 f X} - v(\varphi, f) = P(\varphi , X) \; .
\end{equation}
Utilizing the two equations above, one can determine the form of $v$, or $V$, to reconstruct the corresponding $P(\varphi, X)$ theory.
Then for our two-field completion, using the obtained $V(\varphi , \beta\chi)$, we promote the Lagrangian scalar to the form
\begin{equation}
    {\cal L}_{\rm DBI} = - \sqrt{1 + \left( \partial \chi \right)^2 + f(\beta\chi) \left( \partial\varphi \right)^2} - V(\varphi , \beta\chi) \; , \qquad
    V(\varphi , \beta\chi) \equiv v(\varphi , f(\beta\chi)) \; ,
\end{equation}
acquiring the corresponding (partial) UV theory. The limit $\beta \to \infty$ properly recovers the single-field $P(\varphi, X)$ model.

\subsection{Comparison with other proposals}
\label{subsec:comparison}

In this subsection we discuss relations to the two-field models studied in the literature as (partial) UV completion of $P(\varphi,X)$ models~\cite{Babichev:2017lrx,Babichev:2018twg,Tolley:2009fg,Elder:2014fea,Mizuno:2019pcm,Solomon:2020viz}.
In the prescription of \cite{Babichev:2017lrx,Babichev:2018twg}, the field space metric is flat. Therefore, after requiring the absence of conical singularities in the field space, there is no parameter that describes the properties of the field space, see eq.~\eqref{fieldspace}. As a result, the mass of the extra field is determined by the form of $P(X)$. For a given form of $P(X)$, there is no parameter controlling the mass of extra field and thus the scale at which the corresponding single-field EFT breaks down. This means that the two-field model of \cite{Babichev:2017lrx,Babichev:2018twg} does not allow for a limit in which the single-field $P(X)$ model is recovered with an arbitrarily high precision.

More general prescription called gelaton scenario was originally proposed in \cite{Tolley:2009fg} and then further developed in \cite{Elder:2014fea}. There is a parameter that controls the mass of the extra field and thus the scale at which the corresponding single-field EFT breaks down. By taking a particular limit, one can therefore recover the single-field $P(\varphi,X)$ model with an arbitrarily high precision. However, in this scenario the field space metric and the potential in the two-field model are simultaneously determined by the form of $P(\varphi,X)$, meaning that the field space metric is less controllable than the prescriptions developed in the present paper.

In \cite{Mizuno:2019pcm}, specialized to a single-field DBI action with or without shift symmetry, an extended version of the gelaton scenario was developed, in which the field space metric is specified independently from the form of the single-field action~\footnote{General $P(\varphi,X)$ models were also studied in \cite{Mizuno:2019pcm} but with the original gelaton scenario that we have explained above. On the other hand, \cite{Solomon:2020viz} adopted the original gelaton scenario for all single-field models including the DBI model.}. In particular the field space is specified to the hyperbolic one. The curvature of the hyperbolic field space metric controls the mass of extra field and thus the scale at which the corresponding single-field EFT breaks down. In the limit where the curvature of the field space is infinite, the single-field DBI action with or without shift symmetry is recovered with an arbitrarily high precision.

In the present paper we have developed two different prescriptions of two-field completion: one with linear kinetic terms in subsection \ref{subsec:linear} and the other with the DBI-type kinetic terms in subsection \ref{subsec:DBI}. The first prescription can be considered as a direct generalization of the extended gelaton scenario developed in \cite{Mizuno:2019pcm} to the general $P(\varphi,X)$ models. It can be applied to any $P(\varphi,X)$ that is convex (or concave) with respect to $X$, i.e. $P_{XX}>0$ (or $P_{XX}<0$). The field space metric can be specified to the form \eqref{fieldspace} with an arbitrary positive and non-constant function $f$, independently from the form of $P(\varphi,X)$. The parameter $\beta$ parameterizing the curvature of the field space then controls the mass of the extra field and thus the scale at which the single-field EFT breaks down. In the limit where the curvature of the field space is infinite, the single-field $P(\varphi,X)$ model is recovered with an arbitrarily high precision.

The second prescription developed in the present paper utilizes a two-field system with DBI-type kinetic terms. This is motivated by the two facts:  DBI scalars are free from simple wave caustics~\cite{Mukohyama:2016ipl}; and DBI scalars exist in string theory~\cite{Polchinski:1998rr}. The field space metric (corresponding to the metric of extra dimensions in which the D-brane moves) can be specified to the form \eqref{fieldspace} with an arbitrary positive and non-constant function $f$, independently from the form of $P(\varphi,X)$. The parameter $\beta$ parameterizing the curvature of the field space then controls the mass of the extra field and thus the scale at which the single-field EFT breaks down. In the limit where the curvature of the field space is infinite, the single-field $P(\varphi,X)$ model is recovered with an arbitrarily high precision.

\section{Caustic avoidance}
\label{sec:planar}

As shown in \cite{Babichev:2016hys,Mukohyama:2016ipl}, models of $k$-essence theory in general run into caustics formation within a finite time in the Minkowski spacetime, with a planar-symmetric configuration of the scalar field $\varphi$. The implication of this nature is that such a $k$-essence model should be interpreted as an effective theory, and that a more complete theory needs to take over before the formation of caustics.
This section is devoted to the demonstration of the resolution of a caustics singularity that exists in the single-field effective theory of the type \eqref{LagEFT}, by promoting it to a two-field theory of the type \eqref{LagTwo} that is valid beyond the would-be caustics.

In \cite{Babichev:2016hys,Mukohyama:2016ipl}, the existence of caustics in $k$-essence is shown analytically. For our two-field completion in this paper, we employ numerical integrations, due to the nature of highly nonlinear, coupled partial differential equations in the considered model.
For a concrete demonstration, we make use of the same example single-field model as in \cite{Babichev:2016hys,Mukohyama:2016ipl}, namely a $k$-essence with
\begin{equation}
    P(X) = X + \frac{c}{2} X^2 \; ,
    \label{PX_example}
\end{equation}
where $c$ is a constant. This choice corresponds to $c_0 = 0$, $c_1 = 1$, $c_2 = c/2$ and all other $c_n = 0$ in \eqref{PX_full}.
Using \eqref{V_particular}, this model can be mapped to the choice of $V(\beta \chi)$ in \eqref{LagTwo} as
\begin{equation}
    V = \frac{1}{2 c} \left[ f(\tilde{\chi}) - 1 \right]^2 \; ,\quad \tilde{\chi} = \beta\chi \; ,
    \label{V_example}
\end{equation}
for a given $f(\tilde{\chi})$.
Indeed, eq.~\eqref{chiX_EFT} together with this $V(\beta\chi)=v(f)$ gives the relations $f = 1 + c X$ and $V = c X^2 / 2$, and eq.~\eqref{LagEFT} recovers \eqref{PX_example} as a single-field effective model.

\subsection{Planar symmetric configuration}

Throughout this section we take a planar-symmetric configuration in a Minkowski spacetime, i.e.~without loss of generality,
\begin{equation}
    \varphi = \varphi(t, x) \; , \qquad
    \chi = \chi(t,x) \; ,
\end{equation}
where $t$ and $x$ denote the temporal and one spatial directions, respectively.
With the flat metric, the equations of motion \eqref{EOMTwo_chi} and \eqref{EOMTwo_phi} respectively reduce to
\begin{align}
    \partial_t^2 \chi - \partial_x^2 \chi + \beta \left[ V' - \frac{1}{2}\left( \tau^2 - \zeta^2 \right) f' \right] & = 0 \; ,
    \label{EOM_example_chi}\\
    \partial_t \tau - \partial_x \zeta + \beta \left( \tau \, \partial_t \chi - \zeta \, \partial_x \chi \right) \frac{f'}{f} &= 0 \; ,
    \label{EOM_example_tau}
\end{align}
where we have defined $\tau \equiv \partial_t \varphi$ and $\zeta \equiv \partial_x \varphi$, which suffice to represent the degrees of freedom of the shift-symmetric field $\varphi$.
These two equations, together with the integrability condition
\begin{equation}
    \partial_t \zeta = \partial_x \tau \; ,
    \label{EOM_example_zeta}
\end{equation}
closes the system of equations.
Note that $X = ( \tau^2 - \zeta^2 ) / 2$ in the planar-symmetric configuration.

The one-field reduction of the above model in the limit $\beta \to \infty$ is the $k$-essence \eqref{PX_example}. \
The constraint equation \eqref{chiX_EFT} in this case reads
\begin{equation}
    f = 1 + c X \; .
    \label{const_example}
\end{equation}
Then the only propagating degree of freedom is $\varphi$, and its equation of motion yields
\begin{equation}
    \partial_t \tau
    - \frac{4 \, c \, \tau \zeta}{2 + c \left( 3 \tau^2 - \zeta^2 \right)} \, \partial_x \tau
    - \frac{2 + c \left( \tau^2 - 3 \zeta^2 \right)}{2 + c \left( 3 \tau^2 - \zeta^2 \right)} \, \partial_x \zeta = 0 \; , \qquad
    \partial_t \zeta - \partial_x \tau = 0 \; ,
    \label{EOMEFT_example}
\end{equation}
where $\chi$ has been integrated out thanks to the use of \eqref{const_example}.
These equations \eqref{EOMEFT_example} for $\tau$ and $\zeta$ indeed exactly match the ones obtained starting from the $P(X)$ model in \eqref{PX_example}.

\subsection{Setup of numerical calculation}

We now describe the setup of numerical integrations both for the single-field EFT and for the two-field completion. From here on, $c=1$ is taken for a computational purpose.
Also, as a representative (partial) UV completion, we consider a hyperboloidal field space by taking
\begin{equation}
    f(\tilde\chi) = {\rm e}^{2\tilde\chi} \; , \qquad \tilde\chi \equiv \beta \chi\; .
    \label{f_example}
\end{equation}
We first solve the system of equations \eqref{EOMEFT_example} for $\tau$ and $\zeta$ to observe the formation of caustics in the single-field system, and then compare that to the solutions of the two-field system \eqref{EOM_example_chi}, \eqref{EOM_example_tau} and \eqref{EOM_example_zeta} to show the singularity resolution by invoking the motion of the second field $\chi$.

We choose to have our demonstration go along the example given in \cite{Mukohyama:2016ipl}. To this end, we set the initial condition at $t = 0$ as follows: the reduced single-field case \eqref{EOMEFT_example} admits a class of solutions that obey the equation of differentials in terms of the variables $X=(\tau^2 - \zeta^2)/2$ and $v \equiv - \zeta/\tau$ \cite{Mukohyama:2016ipl},%
\footnote{In terms of the notations used in \cite{Mukohyama:2016ipl}, the solutions of this class are along the $C_-$ characteristics, together with a constant Riemann invariant $\Gamma_-$, defined in eq.~(20) of \cite{Mukohyama:2016ipl}.}
\begin{align}
    \frac{\dd X}{X c_s(X)} - \frac{2 \, \dd v}{1 - v^2} = 0 \; , \quad \mbox{at}\ t=0\; ,
    \label{diff_X_example}
\end{align}
where $c_s$ is defined by and given as
\begin{align}
    c_s \equiv \sqrt{\frac{P_X}{2 X P_{XX} + P_X}} = \sqrt{\frac{1 + X}{1 + 3 X}} \; .
    \label{cs_example}
\end{align}
At time $t=0$, we set $v = 0.8 \exp(- x^2)$ and solve the nonlinear ordinary differential equation \eqref{diff_X_example} for $X$ in one spatial dimension, with the boundary condition fixing $X = 2$ at the boundaries. The exact locations of the spatial boundaries, which we take $x=-10$ and $x=20$ for the numerical computation, are not important as long as they are sufficiently away from the origin $x=0$.
We take the $\tau>0$ branch of this solution as the initial condition for the time evolution. For all the solutions shown in this section, we set the size of each spatial increment to be $\Delta x = 4 \times 10^{-4}$ and time step to be $\Delta t = 2 \times 10^{-4}$ and take the periodic boundary condition for each variable. The numerical methods were implemented in two independent codes, one in C and and the other in Python (with FEniCS package \cite{AlnaesBlechta2015a,LoggMardalEtAl2012a}), and the results have been cross-confirmed.

\subsection{Results}

\begin{figure}[t]
\hfill
\includegraphics[width=0.49\textwidth]{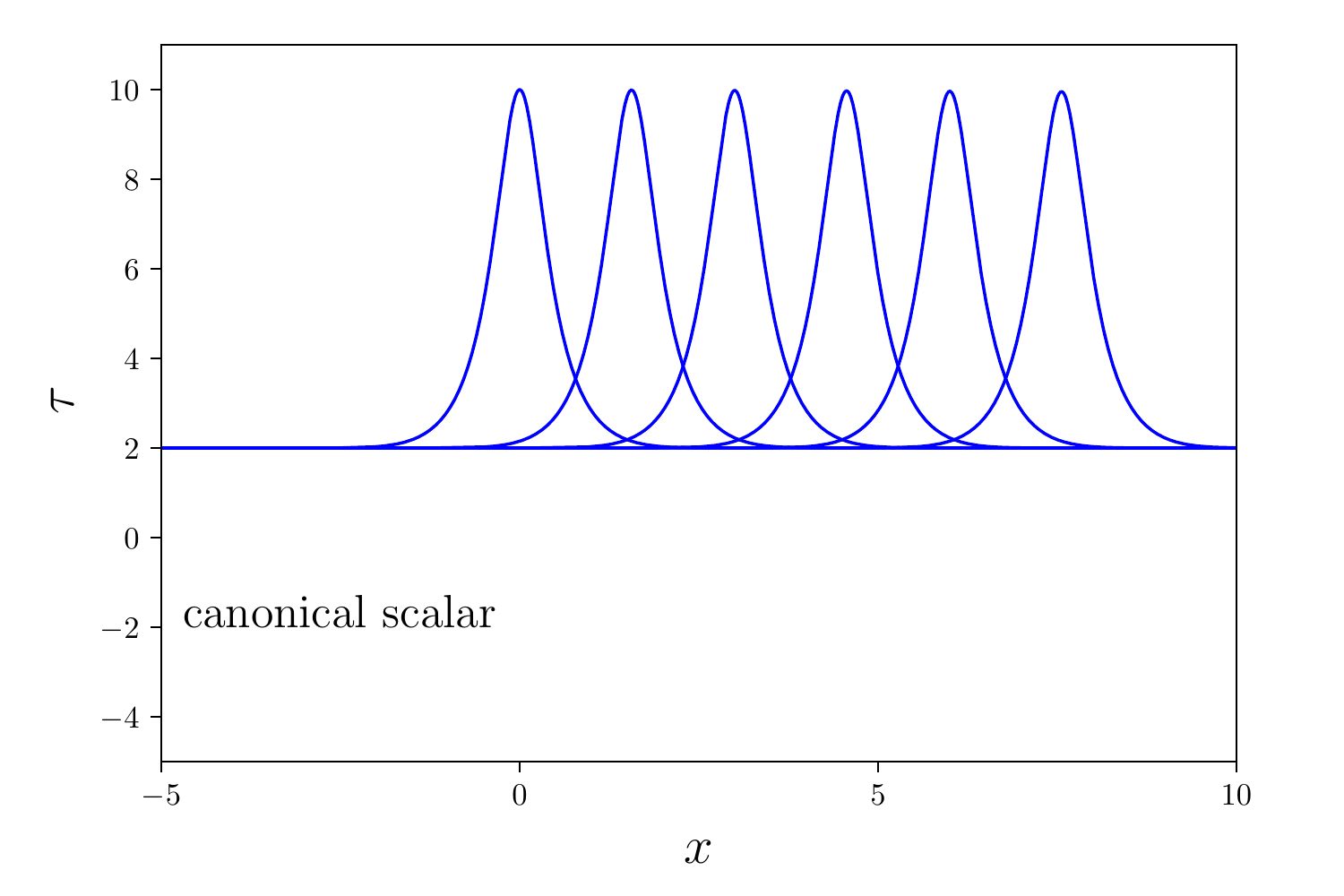}
\includegraphics[width=0.49\textwidth]{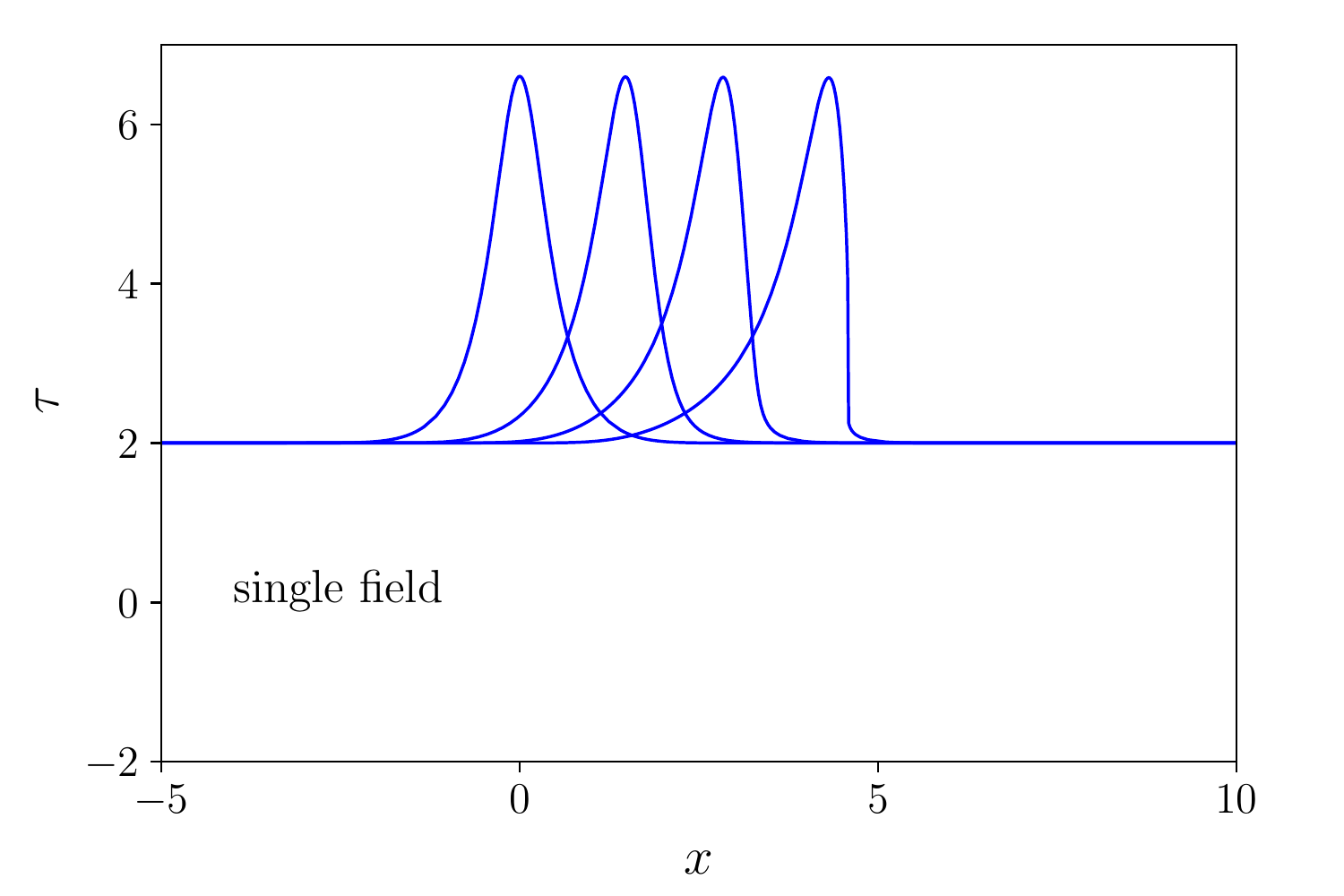}

\includegraphics[width=0.49\textwidth]{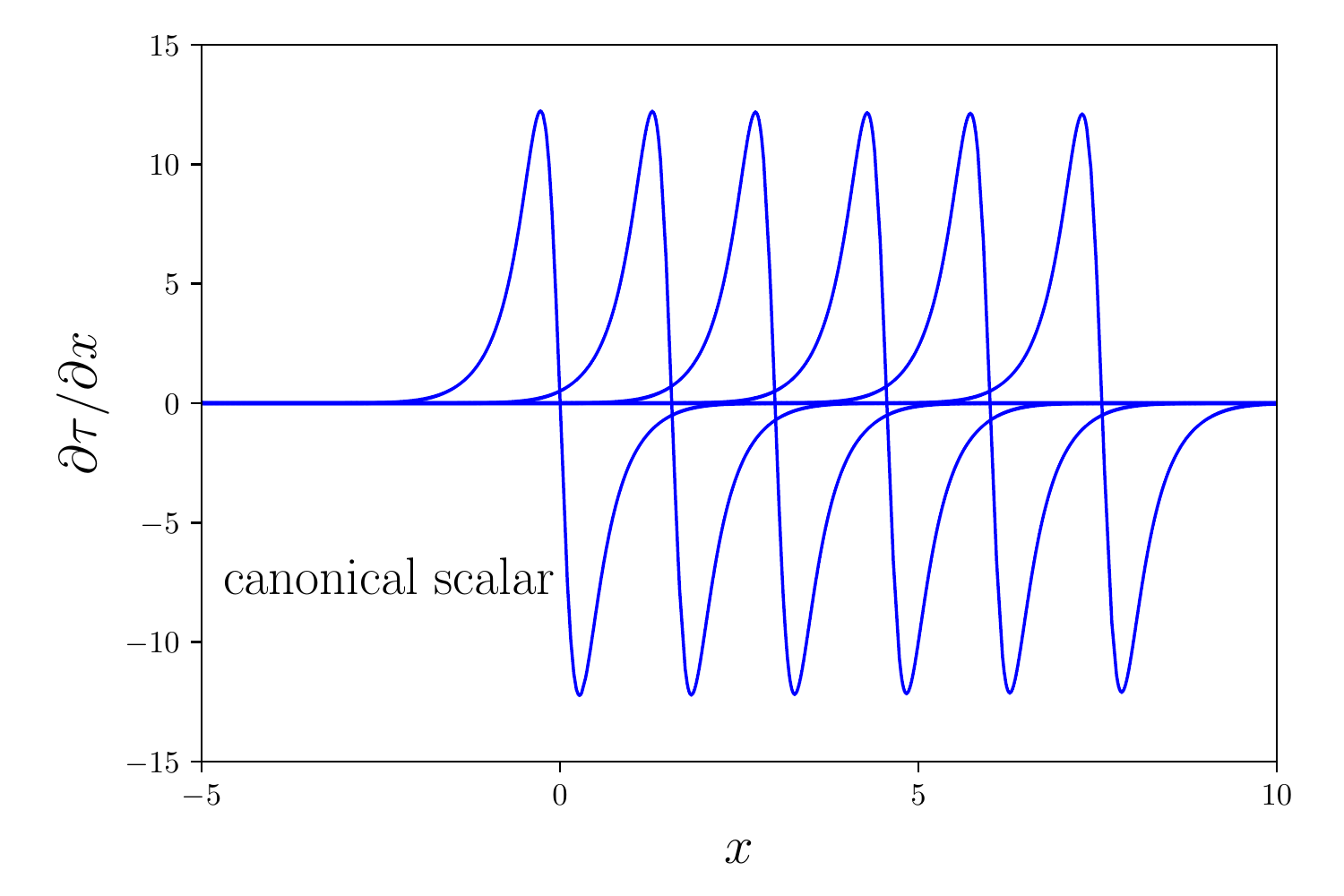}
\includegraphics[width=0.49\textwidth]{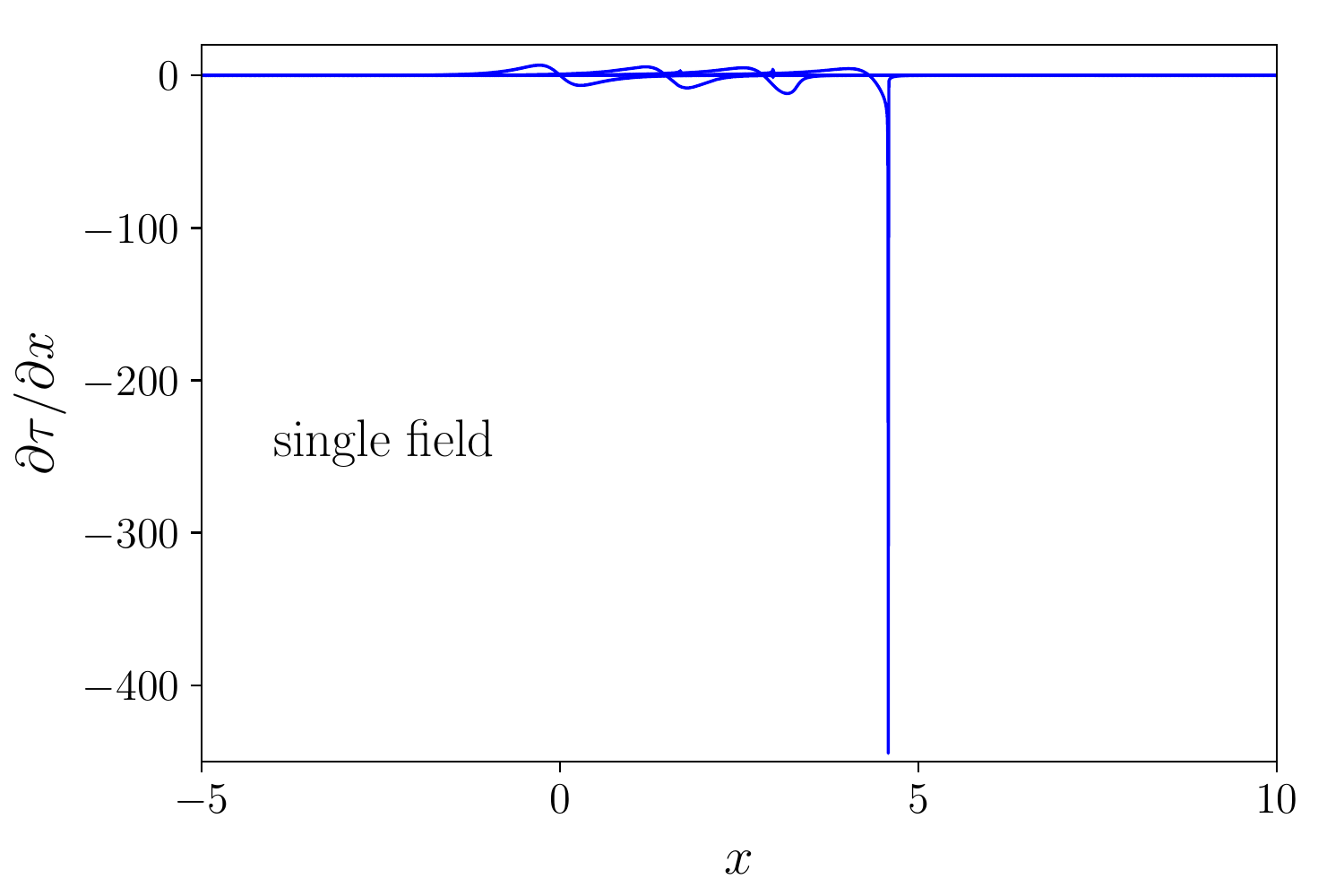} \hfill

\caption{Numerical solutions for single-scalar field models. The right panels show the result for the $P(X) = X + X^2/2$ model \eqref{PX_example}, as compared to the standard canonical case ${\cal L} = X$ on the left panels. The top panels depict the propagation of $\tau$ in each model, and the bottom ones that of $\partial \tau / \partial x$. In each panel, the wave travels to the right, and the snapshots of the wave are taken, from left to right, at $t=0, \, 1.5, \, 3, \, 4.5, \, 6, \, 7.5$ for the left panels, and at $t=0, \, 1.5, \, 3,  \, 4.5$ for the right panels.
Clearly, the wave propagates trivially for the standard canonical scalar, and on the other hand, formation of caustics can be seen near $t=4.5$ in the case of the $P(X)$ model.}
\label{fig:singlefield}
\end{figure}
%
In Fig.~\ref{fig:singlefield}, the numerical solution of the single-field EFT \eqref{EOMEFT_example} is shown on the right panels, while the case of the standard canonical scalar field ${\cal L} = X$ is on the left panels as a reference point. The height of the wave at the initial time differs between the two cases, because the initial condition is obtained using \eqref{diff_X_example} for each case, i.e.~$c_s=1$ for the canonical scalar and $c_s$ given in \eqref{cs_example} for the $k$-essence, with the same $v$ and the same boundary condition for $X$.
In the case of the canonical scalar, the wave simply travels without any change indefinitely. On the other hand, the $k$-essence wave gets distorted while traveling. Its shape changes as if it would fall over in the direction of the propagation. Consequently, the derivative of $\tau$, i.e.~second derivative of $\varphi$, increases over time, and it  becomes divergent around $t=4.5$, as observed in the right bottom panel of Fig.~\ref{fig:singlefield}. The numerical evolution is stopped at this point, beyond which the result could not be trusted. This divergence in the second derivative of the field $\varphi$ is interpreted as the formation of caustics in the considered $k$-essence model.

\begin{figure}[t]
    \centering
    \includegraphics[width=0.485\textwidth]{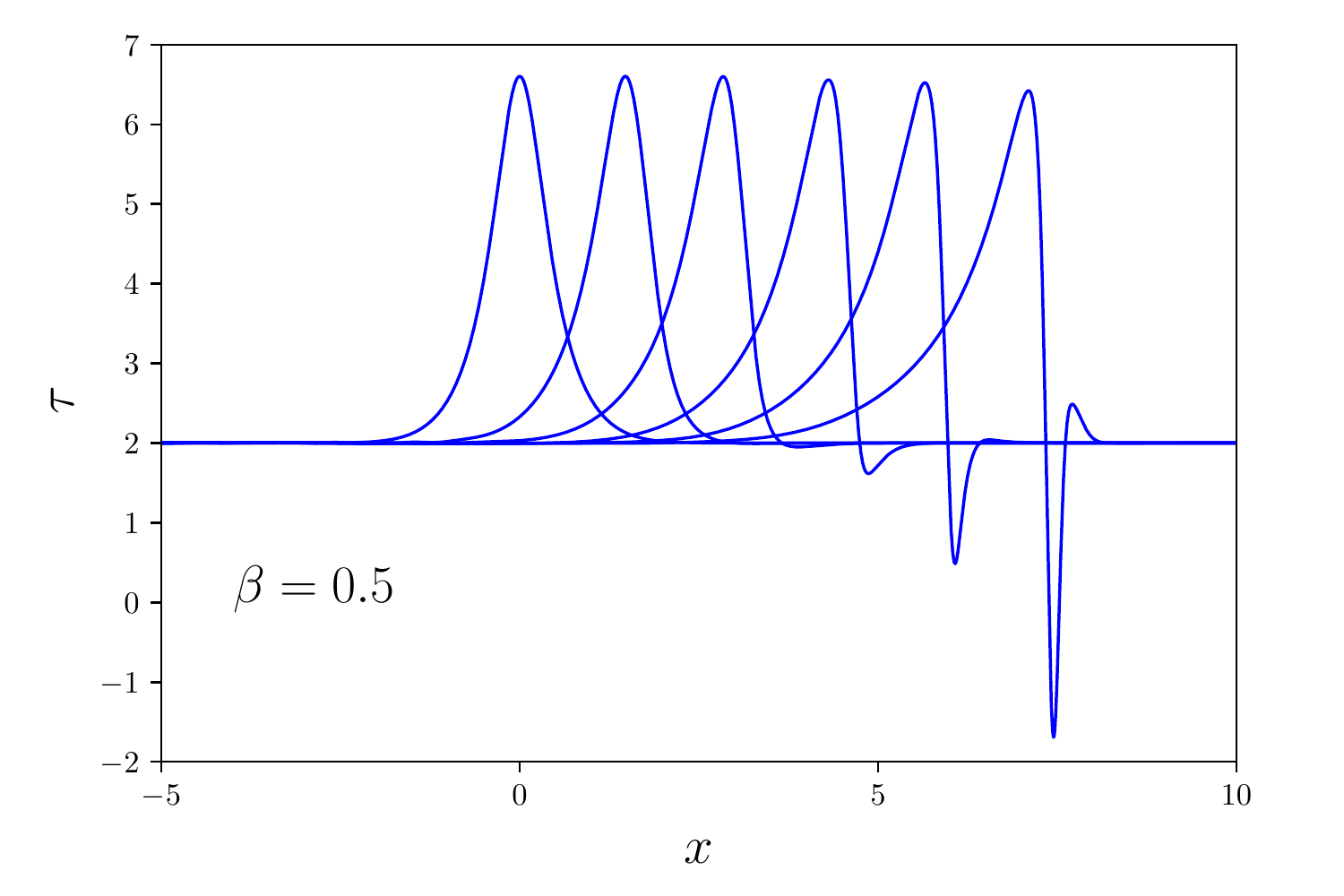} \hspace{0.5mm}
    \includegraphics[width=0.485\textwidth]{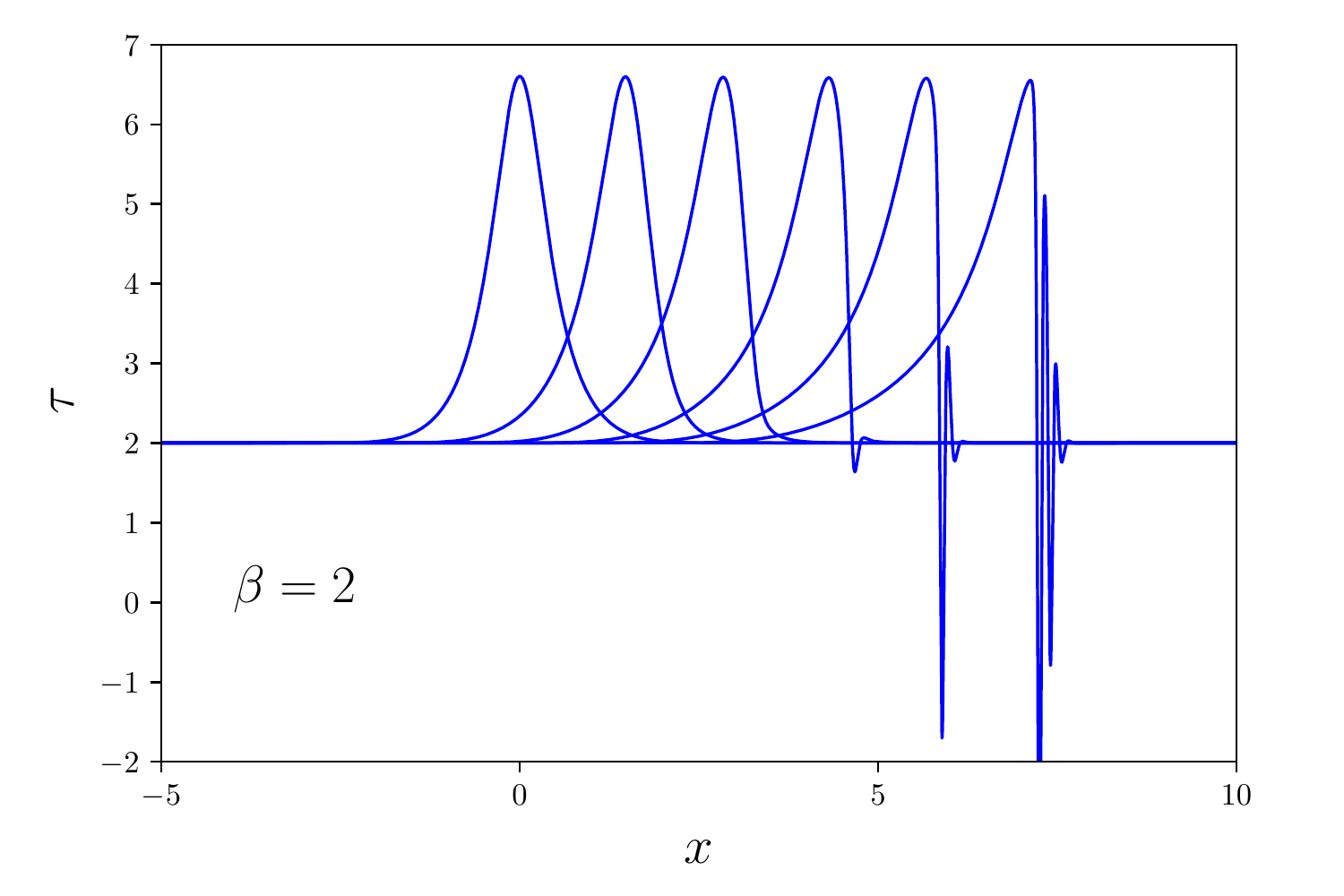}

    \includegraphics[width=0.49\textwidth]{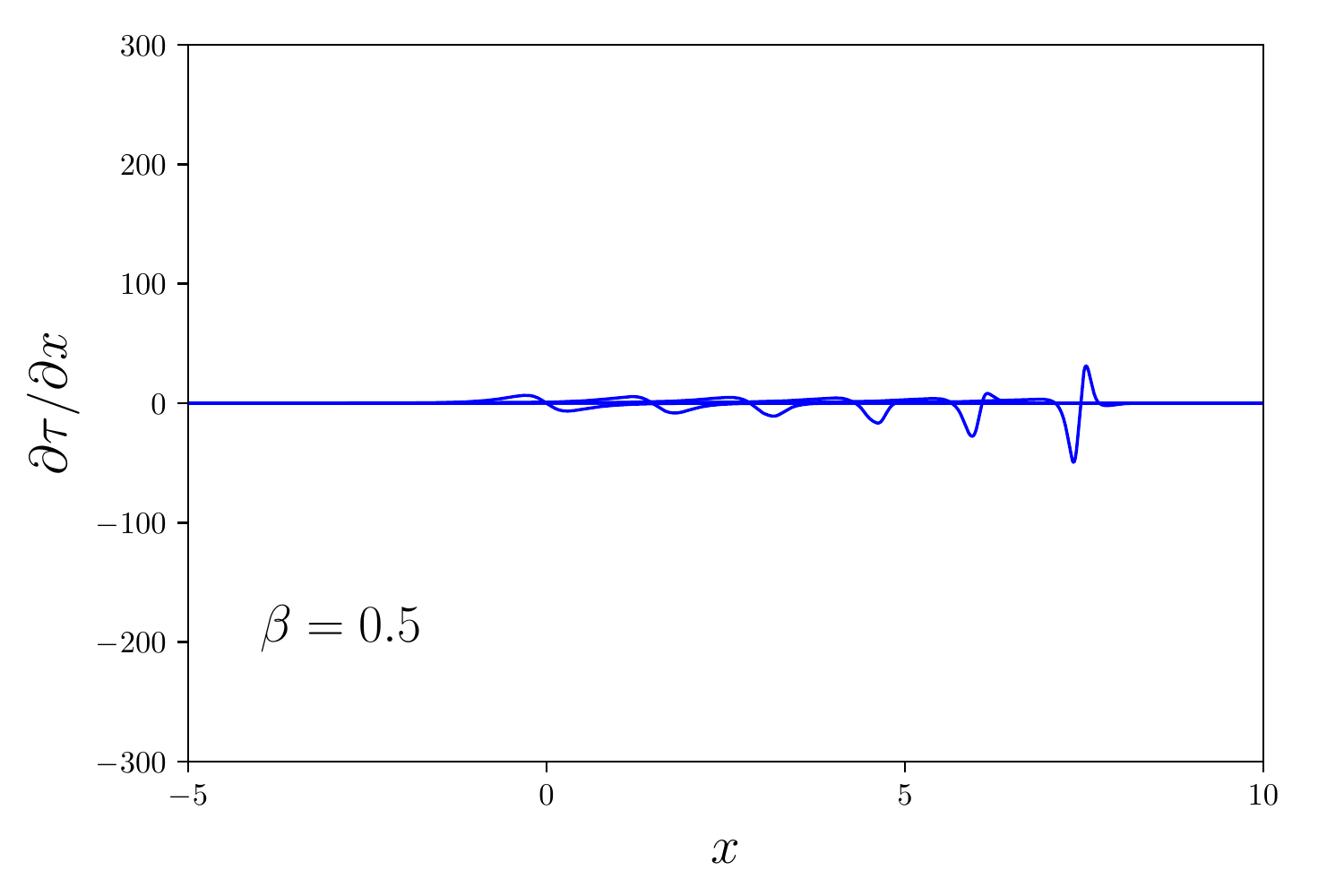}
    \includegraphics[width=0.49\textwidth]{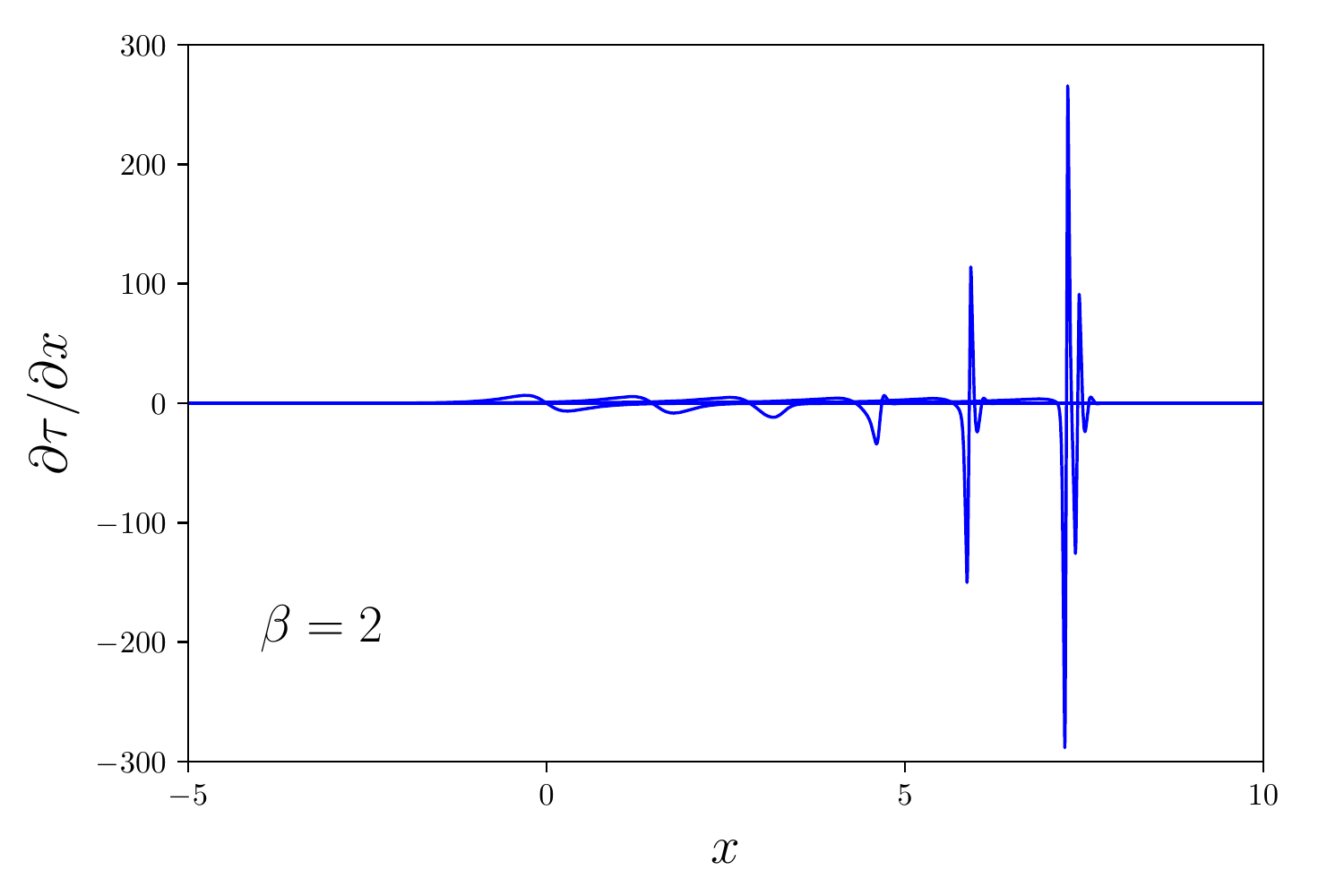} \hfill
    \caption{Numerical solutions for the two-field completion model \eqref{LagTwo} with $f$ and $V$ given by \eqref{f_example} and \eqref{V_example}, respectively, with $c=1$, for two different values of $\beta$. The left panels show the solutions for, from top to bottom, $\tau$ and $\partial \tau / \partial x$ in the case of $\beta = 0.5$, while the right panels are those for $\beta = 2$. The initial conditions are taken to correspond to Fig.~\ref{fig:singlefield}. In each panel, the wave travels to the right, and the snapshots are taken, from left to right, at $t=0, \, 1.5 , \, 3 , \, 4.5 , \, 6 , \, 7.5$. The caustics singularity appearing in the right panels of Fig.~\ref{fig:singlefield} is resolved, more smoothly for a smaller mass scale $\beta = 0.5$ than for a larger one $\beta = 2$.}
    \label{fig:twofields}
\end{figure}
%
Fig.~\ref{fig:twofields} shows the avoidance of the caustics formation in the two-field model \eqref{LagTwo} with $f$ and $V$ given in \eqref{f_example} and \eqref{V_example}, respectively. Here we use $\{ \tau , \, \zeta , \, \tilde\chi \}$ as the variables and solve \eqref{EOM_example_chi}, \eqref{EOM_example_tau} and \eqref{EOM_example_zeta} for the numerics. The same initial conditions are taken for $\tau$ and $\zeta$ as in the single-field case. For the initial conditions for $\chi$ and $\partial_t \chi$, we assume that $\chi$ is stabilized such that the EFT constraint equation \eqref{const_example} is respected both for its value and for its derivative. In particular, the current case with $c=1$ yields to satisfy at the initial time,
\begin{align}
    \tilde\chi = \frac{1}{2} \ln \left( 1 + X \right) \; , \qquad
    \partial_t \tilde\chi
    = \frac{X \left[ \left( 1 - v^2 \right)\left( 1 + X \right) + 2 v c_s X \right] \partial_x v}{\left( 1 - v^2 \right)\left( 1 + X \right) \left[ \left( 1 + X \right) v^2 - 1 - 3 X \right]} \; , \quad \mbox{at}\ t=0 \; ,
\end{align}
where the time derivatives of $\tau$ and $\zeta$ have been replaced by using the equations of motion on the constrained hypersurface, i.e.~\eqref{EOMEFT_example}, and the spatial derivative of $X$ is replaced by \eqref{diff_X_example}. Using \eqref{cs_example} for $c_s$, taking $v = 0.8 \exp(-x^2)$ and fixing $X$ at the initial time as explained around \eqref{diff_X_example}, the above equations uniquely determine the initial conditions for $\tilde\chi$ and $\partial_t \tilde\chi$.
We again take the periodic boundary condition.

Comparing Figs.~\ref{fig:singlefield} and \ref{fig:twofields}, it is evident that the two-field case is free from the divergence in $\partial\tau / \partial x$, which is a second derivative of the field $\varphi$, appearing around $t=4.5$, implying that the two-field completion indeed removes the caustic singularity that appears present in its low-energy single-field EFT.
The single-field EFT well describes the evolution of the more fundamental, underlying system until it is about to evolve into the caustics. It is the parameter $\beta$ that controls the scale at which the effect of the second field $\chi$ starts playing a role. As is seen from \eqref{EOM_example_chi}, $\beta$ corresponds to the mass scale of $\chi$, and the larger the value of $\beta$, the larger the mass. Thus, for a smaller value of $\beta$, the single-field EFT breaks down at a lower energy scale, i.e.~at an earlier stage of a caustic formation. This expectation is verified by observing that the $\beta=0.5$ case starts deviating from the single-field EFT dynamics already around $t=3$, while the deviation starts occurring only around $t=4.5$ for $\beta=2$. As a result, the former goes through a smoother evolution than the latter, which carries sharper peaks both in $\tau$ and $\partial \tau / \partial x$ (and other variables as well).

\begin{figure}[t]
    \centering
    \includegraphics[width=0.55\textwidth]{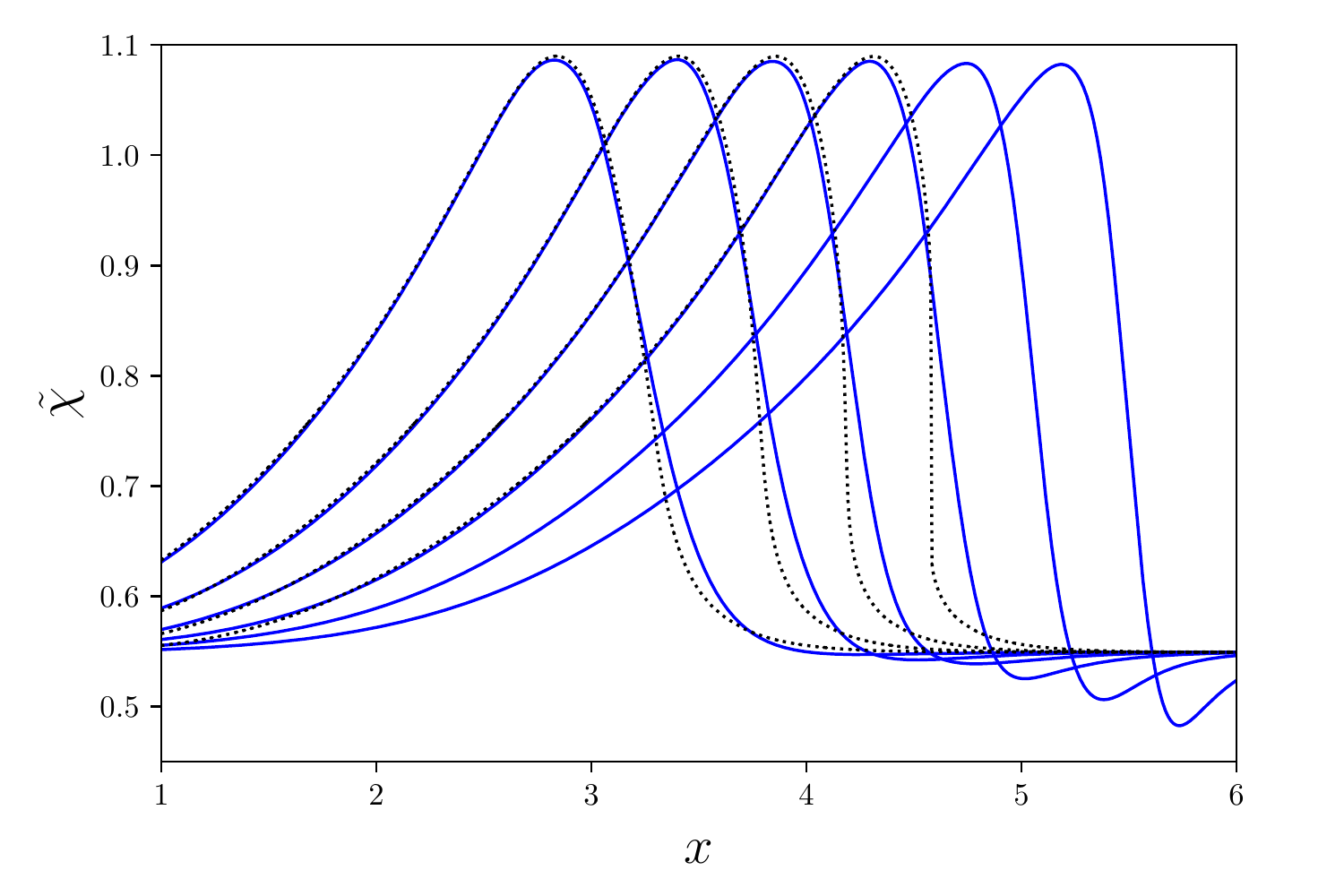}
    \caption{Comparison between $\tilde\chi = \beta \chi$ in the two-field system with $\beta = 0.5$ (blue solid curve) and the value of $\ln (1 + X) / 2$ in the single-field system (black dotted) around the time of caustic formation. These two quantities coincide in the limit $\beta \to \infty$ through the constraint equation \eqref{const_example} with $f$ given in \eqref{f_example}, and in this sense, this plot indicates the deviation from the single-field EFT. The snapshots of the waves are taken at $t= 3, \, 3.5 , \, 4 , \, 4.5 , \, 5 , \, 5.5$, although the evolution of the single-field model is stopped at $t=4.5$.}
    \label{fig:chit_compare}
\end{figure}
\begin{figure}[th]
    \centering
    \includegraphics[width=0.55\textwidth]{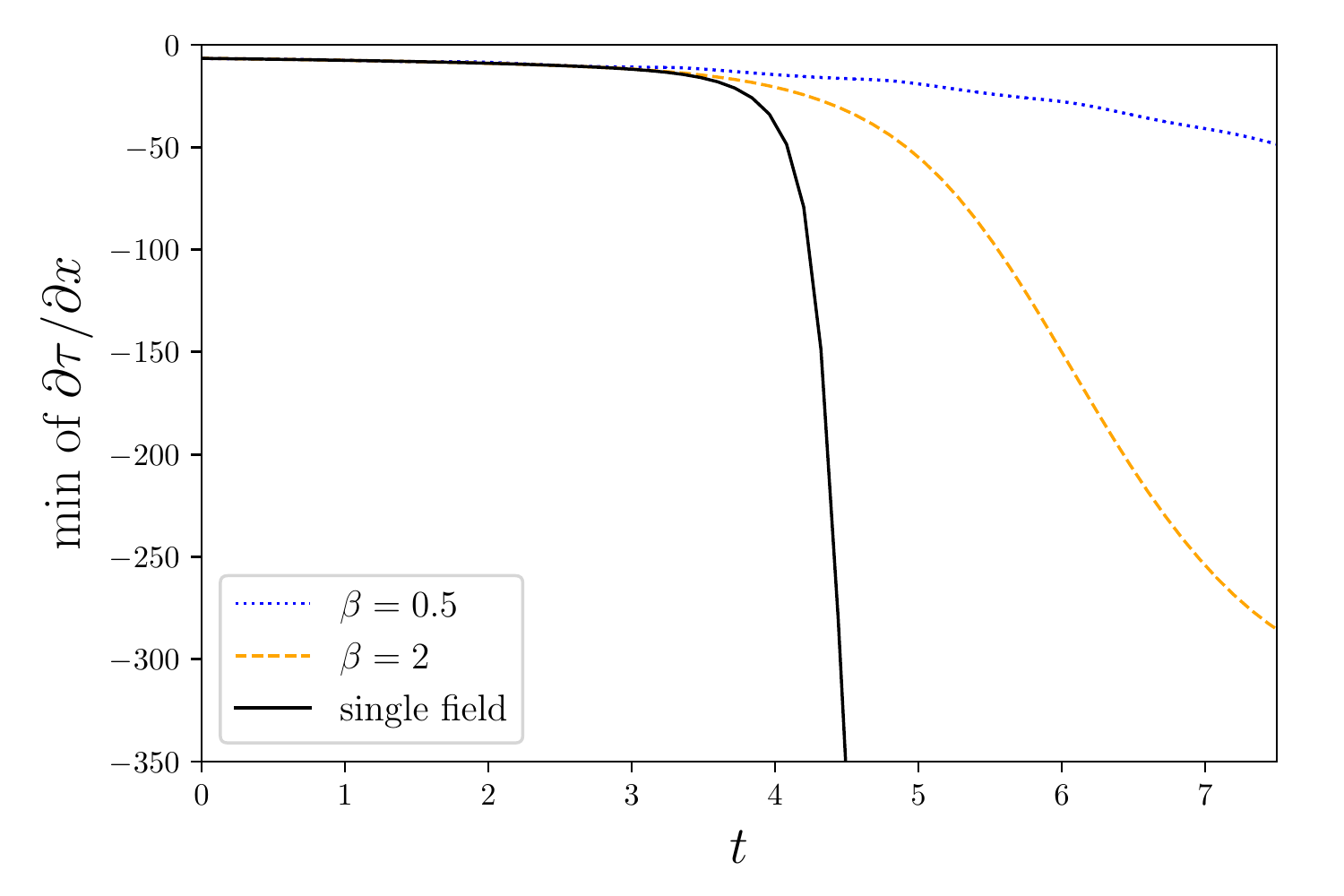}
    \caption{The time evolution of the minimum values of $\partial \tau / \partial x$ (correspondingly to the largest values of $|\partial \tau / \partial x|$), for the single-field and the two-field cases. Caustics forms in the single-field $k$-essence (black solid curve) around $t=4.5$, and it is ameliorated in the two-field completion (blue dotted and orange dashed curves respectively for $\beta=0.5$ and $\beta=2$). }
    \label{fig:dtaudx_compare}
\end{figure}
%
In the EFT limit $\beta \to \infty$, the constraint equation \eqref{chiX_EFT}, or \eqref{const_example}, should be fully respected. In other words, the deviation of $\tilde\chi$ from the value of $\ln ( 1 + X ) / 2$ is an indicator of the departure from the EFT. Fig.~\ref{fig:chit_compare} compares $\tilde\chi$ in the two-field model with $\beta = 0.5$ and $\ln ( 1 + X ) / 2$ computed from the numerical result of the single-field case. The caustics would form around $t = 4.5$ for the latter, and indeed the deviation increases toward this moment. While the shape of the wave appears to fall over in the direction of the propagation in the single-field case, it is smoothed out in the two-field case. The presence of this second field $\chi$ is crucial for the resolution of the caustic singularity.
Finally, Fig.~\ref{fig:dtaudx_compare} shows the time evolution of the minimum values (i.e.~the largest amplitudes) of $\partial \tau / \partial x$ to compare the cases of single field, two fields with $\beta = 0.5$ and two fields with $\beta = 2$. This clearly depicts the divergence in the single-field EFT, while the values of $\partial \tau / \partial x$ are well under control in the two-field completed model. It is again shown that the $\beta = 0.5$ case renders the caustics harmless more efficiently than $\beta = 2$.
To our knowledge, this is the first numerical presentation of the caustic formation in a $k$-essence model and of its resolution in its UV completed version.

To summarize, the two-field completion indeed removes the caustics, and such a partial UV completion is necessary to describe the evolution of a physical system close to and beyond the (would-be) singularity. The parameter $\beta$ control the mass/energy scale of the EFT breakdown, and essentially a measure of the onset of the UV physics.
Hence it is a natural question to ask and is of interest to investigate what and how much influence the UV effects produce on physical systems.
In the next section, we therefore explore and analyze the two-field system in the cosmological settings.

\section{Cosmological applications}
\label{sec:cosmology}

In this section, we analyze the two-field completion model of $k$-essence theory proposed in Sec.~\ref{sec:twofield}, aiming for cosmological applications. For the purpose of computational ease and of intuitive illustration, we focus our detailed analysis on the case of linear kinetic terms presented in subsection~\ref{subsec:linear}.
Using the two-field model with linear kinetic terms that is minimally coupled to gravity, the full action of our interest is
\begin{equation}
    S = \int \dd^4x \sqrt{-g} \left[ \frac{\Mpl^2}{2} R - \frac{1}{2} \left( \partial\varphi \right)^2 - \frac{f(\beta\chi)}{2} \left( \partial\varphi \right)^2 - V(\beta\chi) \right] \; ,
    \label{action}
\end{equation}
where $\Mpl$ is the reduced Planck mass and $R$ is the Ricci scalar associated with the spacetime metric.
In the following subsections, we first consider the flat Friedmann-Lema\^{i}tre-Robertson-Walker (FLRW) background and then proceed to the perturbations around it.
In view of the cosmological application, we keep including gravity throughout our analysis, which is a secure improvement compared to the one in \cite{Babichev:2018twg}.

\subsection{FLRW background}
\label{subsec:FLRW}

For the cosmological background, we take the flat FLRW metric
\begin{equation}
    \dd s^2 = - N^2(t) \, \dd t^2 + a^2(t) \, \delta_{ij} \, \dd x^i \dd x^j \; ,
    \label{FLRW}
\end{equation}
and the homogeneous modes for the fields
\begin{equation}
    \langle \varphi \rangle = \phi(t) \; , \qquad
    \langle \chi \rangle = \chi(t) \; ,
    \label{background}
\end{equation}
with some abuse of notation.
Here $a(t)$ is the scale factor and $N(t)$ is the lapse function, which we will later set to unity.
Then the background action of \eqref{action} reads
\begin{equation}
S^{(0)} = {\cal V} \int N \dd t \, a^3 \left[ - 3 \Mpl^2 \, \frac{\left( \partial_t a \right)^2}{a^2 N^2}
+ \frac{1}{2} \, \frac{\left( \partial_t \chi \right)^2}{N^2}
+ \frac{f}{2} \, \frac{\left( \partial_t \phi \right)^2}{N^2}
- V \right] \; ,
\end{equation}
where ${\cal V}$ is the comoving spatial $3$-volume, and hereafter $f$ and $V$ denote the background values of the corresponding functions, i.e.~$f = f(\beta\chi(t))$ and $V = V(\beta \chi(t))$, respectively.
The background dynamics is governed by the equations of motion
\begin{align}
& \frac{\partial_t^2 \chi}{N^2} + \left( 3 H - \frac{\partial_t N}{N^2} \right) \frac{\partial_t \chi}{N}
+ \beta \left[ V' - \frac{f'}{2} \, \frac{\left( \partial_t \phi \right)^2}{N^2} \right]= 0 \; ,
\label{EOMback_chi}\\
& \frac{\partial_t^2 \phi}{N^2} + \left( 3 H - \frac{\partial_t N}{N^2}
+  \beta \, \frac{f'}{f} \, \frac{\partial_t \chi}{N} \right) \frac{\partial_t \phi}{N} = 0 \; ,
\label{EOMback_phi}
\end{align}
where prime on $f$ and $V$ denotes derivative with respective their argument, together with the Friedmann equation
\begin{equation}
\begin{aligned}
3 \Mpl^2 H^2 & = \frac{f}{2} \, \frac{\left( \partial_t \phi \right)^2}{N^2}
+ \frac{1}{2} \, \frac{\left( \partial_t \chi \right)^2}{N^2} + V \; ,
\end{aligned}
\label{Friedmann}
\end{equation}
where $H \equiv \partial_t a / (aN)$ is the Hubble expansion rate. The above three equations close the system.
The background energy density $\bar\rho$ and the pressure $\bar p$ read
\begin{align}
\bar\rho &
= \frac{f}{2} \, \frac{\left( \partial_t \phi \right)^2}{N^2}
+ \frac{1}{2} \, \frac{\left( \partial_t \chi \right)^2}{N^2} + V \; , \qquad
\bar p
= \frac{f}{2} \, \frac{\left( \partial_t \phi \right)^2}{N^2}
+ \frac{1}{2} \, \frac{\left( \partial_t \chi \right)^2}{N^2} - V \; .
\label{rho_BG}
\end{align}
So far the above equations are for the full two-field system.

We now proceed to the EFT reduction of the full system to a low-energy single-field one and then to the leading-order correction for $\beta \gg 1$. From here on, we choose to set $N=1$.
For a consistent expansion for large $\beta$, we expand as
\begin{equation}
\chi = \epsilon \chi_1 + \epsilon^{2} \chi_2 + \epsilon^3 \chi_3 + \dots \; , \qquad
\phi = \phi_0 + \epsilon \phi_1 + \epsilon^{2} \phi_2 + \dots \; , \qquad
H = H_0 + \epsilon H_1 + \epsilon^2 H_2 + \dots \; ,
\label{back_expansion}
\end{equation}
where $\epsilon$ is the expansion parameter, with $\beta = {\cal O}(\epsilon^{-1})$, and subscripts $0,1,2,\dots$ keep track of the expansion order. Note that $\phi$ and $\chi$ start from the $0$th and $1$st orders in $\epsilon$, respectively.
We expand the equations of motion, \eqref{EOMback_chi} and \eqref{EOMback_phi}, and the Friedmann equation \eqref{Friedmann} for small $\epsilon$. We first notice that the equation of motion for $\chi$ starts with ${\cal O}(\epsilon^{-1})$, and the leading order for the rest of the equations is ${\cal O}(\epsilon^0)$. Picking up the leading order of each equation of \eqref{EOMback_chi}, \eqref{EOMback_phi} and \eqref{Friedmann}, we find
\begin{align}
& V_0' = f_0' \, X_0 \; ,
\label{EOMback_chi_0} \\
& \partial_t^2 \phi_0
+ 3 H_0 \,
c_{s,0}^2 \, \partial_t \phi_0 = 0 \; ,
\label{EOMback_phi_0}\\
& 3 \Mpl^2 H_0^2 = f_0 X_0 + V_0 \; ,
\label{Friedmann_0}
\end{align}
where $V_0 \equiv V(\beta \chi_1)$, $f_0 \equiv f (\beta \chi_1)$, and $X_0 \equiv \left( \partial_t \phi_0 \right)^2 / 2$, and the $0$th-order sound speed $c_{s,0}$ in this model takes the form
\begin{equation}
    c_{s,0}^2 \equiv \frac{\partial_t p_0}{\partial_t \rho_0} = \frac{f f_0' V_0'' - f f_0'' V_0'}{f f_0' V_0'' + \left( 2 f_0'{}^2 - f f_0'' \right) V_0'} \; ,
    \label{cs0}
\end{equation}
where $\rho_0 \equiv f_0 X_0 + V_0$ and $p_0 \equiv f_0 X_0 - V_0$ are the $0$th-order energy density and pressure, respectively.
In deriving \eqref{EOMback_phi_0}, the time derivative of \eqref{EOMback_chi_0} was also imposed.
Recalling the correspondence with the $P(X)$ theory, $P \leftrightarrow f X - V$ and $P_X \leftrightarrow f$, eqs.~\eqref{EOMback_phi_0} and \eqref{Friedmann_0} exactly reproduce the equations for the effective single-field $P(X)$ theory.

To compute the higher orders in small $\beta^{-1}$, let us expand the energy density as
\begin{equation}
    \bar\rho = \rho_0 + \rho_1 + \rho_2 + \dots \; ,
\end{equation}
where $\rho_0$ is given below \eqref{cs0} and
\begin{align}
    \rho_1 & \equiv 6 \Mpl^2 H_0 H_1
    = \frac{f_0}{c_{s,0}^2} \, \partial_t \phi_0 \, \partial_t \phi_1 \; ,
    \label{rho1} \\
    \rho_2 & \equiv 3 \Mpl^2 \left( 2 H_0 H_2 + H_1^2 \right)
    \nonumber\\ &
    = \frac{f_0}{c_{s,0}^2} \, \partial_t \phi_0 \, \partial_t \phi_2
    + \frac{f_0}{6 c_{s,0}^4} \left( 3 + \frac{\partial_t c_{s,0}^2}{c_{s,0}^2 H_0} \right) \left( \partial_t \phi_1 \right)^2
    \nonumber\\ & \quad
    - \frac{f_0^2}{\beta^2 f_0'{}^2 c_{s,0}^2} \left[
    \frac{3 \left( 1 - c_{s,0}^2 \right)^2 f_0 X_0}{\Mpl^2}
    + 3 \left( 1 - c_{s,0}^2\right) H_0 \, \partial_t c_{s,0}^2
    - \frac{9 \left( 1 - c_{s,0}^2 \right)^2}{2}
    \left( 3 \, c_{s,0}^2 + 2 \left( 1 - c_{s,0}^2 \right) \frac{f_0 f_0''}{f_0'{}^2} \right) H_0^2
    \right] \; ,
\end{align}
where the second equality in each equation above comes from the Friedmann equation \eqref{Friedmann}.
The higher orders of $\chi$, i.e.~$\chi_2 , \, \chi_3 , \, \dots$, can be solved iteratively from \eqref{EOMback_chi} with respect to other variables of lower orders,
\begin{align}
    \chi_2 & = \frac{1 - c_{s,0}^2}{2 \beta f_0' X_0} \, \rho_1 \; ,
    \label{chi2} \\
    \chi_3 & = \frac{1 - c_{s,0}^2}{2 \beta f_0' X_0} \, \rho_2
    + \left[ 2 \, \partial_t c_{s,0}^2
    - 3 \left( 1 - c_{s,0}^2 \right)
    \left( 4 \, c_{s,0}^2
    + \left( 1 - c_{s,0}^2 \right) \frac{f_0 f_0''}{f_0'{}^2} \right) H_0
    \right] \frac{\rho_1^2}{48 \beta H_0 f_0 f_0' X_0^2}
    \nonumber \\ & \quad
    - \left[
    2 \, \partial_t c_{s,0}^2
    - 3 \left( 1 - c_{s,0}^2 \right)
    \left( 3 \, c_{s,0}^2 - 1
    + \left( 1 - c_{s,0}^2 \right) \frac{f_0 f_0''}{f_0'{}^2} \right) H_0
    \right]
    \frac{3 \left( 1 - c_{s,0}^2 \right) H_0 f_0^2}{4 \beta^3 f_0'{}^3 X_0}
    - \frac{3 \left( 1 - c_{s,0}^2 \right)^2 f_0^3}{2 \beta^3 \Mpl^2 f_0'{}^3} \; .
    \label{chi3}
\end{align}
Combining with the above equations, the $\phi$'s E.o.M.~\eqref{EOMback_phi} translates to the equations for $\rho_1$ and $\rho_2$, giving
\begin{align}
    & \partial_t \rho_1 + \left[ 3 H_0 \left( 1 + c_{s,0}^2 \right) + \frac{f_0 X_0}{\Mpl^2 H_0} \right] \rho_1 = 0 \; ,
    \label{EOM_rho1}\\
    & \partial_t \rho_2+ \left[ 3 H_0 \left( 1 + c_{s,0}^2 \right)
    + \frac{f_0 X_0}{\Mpl^2 H_0} \right] \rho_2 =
    \left(
    \frac{\partial_t c_{s,0}^2}{4 f_0 X_0}
    - \frac{1 + c_{s,0}^2}{2 \Mpl^2 H_0}
    + \frac{f_0 X_0}{12 \Mpl^4 H_0^3}
    \right) \rho_1^2
    \nonumber\\ & \qquad
    - \frac{9 \left( 1 - c_{s,0}^2 \right) f_0^2 H_0^2}{\beta^2 f_0'{}^2}
    \left[
    \partial_t c_{s,0}^2
    + \frac{\left( 1 - c_{s,0}^2 \right) \left( 3 \left( 1 - 3 \, c_{s,0}^2 \right) \Mpl^2 H_0^2 + 2 f_0 X_0 \right)}{2 \Mpl^2 H_0}
    - 3 \left( 1 - c_{s,0}^2 \right)^2 \frac{f_0 f_0''}{f_0'{}^2} \, H_0
    \right]
    \; .
    \label{EOM_rho2}
\end{align}
We have obtained up to the first $3$ orders of equations.
We note that, at the first order, only the dispersion relation for $\rho_1$ is modified as seen in \eqref{EOM_rho1}, and then \eqref{EOM_rho2} indicates that the lower-order terms act as source for the second and higher orders.
Higher-order equations can be obtained by a straightforward extension of the above methodology.

In this subsection, we have shown that the single-field reduction from the two-field UV theory correctly reproduces the expected $k$-essence as EFT in the limit of large mass of $\chi$ field, $\beta \to \infty$. The corrections to the leading-order EFT can be unambiguously calculated by the method of perturbative expansion, up to an arbitrary order in $\beta^{-1}$. We have so far demonstrated this for the flat FLRW background, and, in the following subsection, we extend the analysis to the cosmological perturbations.

\subsection{Cosmological perturbations}
\label{subsec:perturbations}

In this subsection, we proceed to the perturbations around the FLRW background. It can be trivially seen that the tensor and vector sectors are as standard as in any models of scalar fields minimally coupled to gravity, and thus we look into the details of the scalar sector below.
For the scalar sector, we expand the variables as
\begin{equation}
\varphi(t, \bm{x}) = \phi(t) + \delta\varphi(t, \bm{x}) \; , \qquad
\chi(t, \bm{x}) = \chi(t) + \delta\chi(t, \bm{x}) \; ,
\label{pert_fields}
\end{equation}
for the scalar fields and
\begin{equation}
g_{00}(t, \bm{x}) =
- 1 - 2 \Phi(t, \bm{x}) \; , \quad
g_{0i}(t, \bm{x}) =
a(t) \, \partial_i B(t, \bm{x}) \; , \quad
g_{ij}(t, \bm{x}) = a^2(t) \left[ \left( 1 + 2 \, \Psi (t,\bm{x}) \right) \delta_{ij} + 2 \, \partial_i \partial_j E (t,\bm{x}) \right] \; ,
\label{pert_metric}
\end{equation}
for the metric.
The linear-order action vanishes after using the background equations.
In deriving the quadratic action, we take the spatially flat gauge, namely $\Psi = E = 0$. It is then clear that $\Phi$ and $B$ are non-dynamical variables and can be eliminated by replacing them in terms of the dynamical degrees of freedom. By employing the Faddeev-Jackiw method \cite{Faddeev:1988qp}, we obtain the quadratic action in terms only of the dynamical variables $\delta_i \equiv ( \delta\varphi , \delta\chi )$ as, in the Fourier space,
\begin{equation}
S^{(2)}_{\rm scalar} = \frac{1}{2} \int \dd t \, \dd^3k \left(
\partial_t \delta^\dagger_{i} \, T_{ij} \, \partial_t \delta_j
+ \partial_t \delta^\dagger_{i} \, X_{ij} \, \delta_j
- \delta^\dagger_{i} \, X_{ij} \, \partial_t \delta_j
- \delta_i^\dagger \, \Omega^2_{ij} \, \delta_j
\right) \; ,
\label{action_scalar}
\end{equation}
up to total derivatives. Here, $T$, $X$ and $\Omega^2$ are real, $2 \times 2$ matrices constructed by the background quantities and have the properties $T^T = T$, $X^T = - X$ and $(\Omega^2)^T = \Omega^2$. Their explicit expressions are
\begin{align}
T & = a^3 \left(
\begin{array}{cc}
f & 0 \\ 0 & 1
\end{array}
\right) \; , \qquad
X = a^3 \left(
\begin{array}{cc}
0 & \displaystyle \frac{\beta}{2} \, f' \, \partial_t \phi \\
\displaystyle - \frac{\beta}{2} \, f' \, \partial_t \phi & 0
\end{array}
\right) \; ,
\label{mat_TX} \\
\Omega^2_{11} & =
a^3 f \left[ \frac{k^2}{a^2} + \frac{3 f \left( \partial_t \phi \right)^2}{\Mpl^2} - \frac{f \left( \partial_t \phi \right)^2}{2 \Mpl^4 H^2} \left( f \left( \partial_t \phi \right)^2 + \left( \partial_t \chi \right)^2 \right) \right] \; ,
\label{mat_O11}\\
\Omega^2_{12} & = \Omega^2_{21} = a^3 \left[
- \frac{\beta^2}{2} \left( \frac{f'{}^2}{f}-f'' \right) \partial_t \phi \, \partial_t \chi
+ \frac{\beta f \, \partial_t \phi}{\Mpl^2 H} \, V'
+ \frac{3 f \, \partial_t \phi \, \partial_t \chi}{\Mpl^2}
- \frac{f \, \partial_t \phi \, \partial_t \chi}{2 \Mpl^4 H^2} \left( f \left( \partial_t \phi \right)^2 + \left( \partial_t \chi \right)^2 \right)
\right] \; ,
\label{mat_O12} \\
\Omega^2_{22} & = a^3 \Bigg[
\frac{k^2}{a^2}
+ \beta^2 \left( V'' - \frac{f''}{2} \left( \partial_t \phi \right)^2 \right)
+ \frac{2 \beta \, \partial_t \chi \, V'}{\Mpl^2 H}
+ \frac{3 \left( \partial_t \chi \right)^2}{\Mpl^2}
- \frac{\left( \partial_t \chi \right)^2}{2 \Mpl^4 H^2} \left( f \left( \partial_t \phi \right)^2 + \left( \partial_t \chi \right)^2 \right)
\Bigg] \; .
\label{mat_O22}
\end{align}
Variation of \eqref{action_scalar} with respect to $\delta_i$ provides the equations of motion,
\begin{equation}
\partial_{t}^2 \delta + T^{-1} \left( 2 X + \partial_t T \right) \partial_t \delta + T^{-1} \left( \Omega^2 + \partial_t X \right) \delta = 0 \; ,
\label{EOM_pert}
\end{equation}
in the matrix form. This is so far the genuine two-field system. In what follows, we show that reduction to the one-field EFT is successfully done for large $\beta$ and that higher-order corrections can be iteratively computed with no ambiguity.

In order to expand the system in terms of small $\beta^{-1}$, we decompose the background quantities as in \eqref{back_expansion} and the perturbation variables as
\begin{equation}
\delta\varphi = \delta\varphi_0 + \epsilon \, \delta\varphi_1 + \dots \; , \qquad
\delta\chi = \epsilon \, \delta\chi_1 + \epsilon^2 \delta\chi_2 + \dots \; ,
\label{pert_expansion}
\end{equation}
where $\epsilon = {\cal O}(\beta^{-1})$. Note that, similarly to the background, the massive field $\delta\chi$ in the EFT reduction starts at the order of ${\cal O}(\epsilon)$, while the propagating field $\delta\varphi$ in the EFT starts at ${\cal O}(\epsilon^{0})$, as seen below.
Collecting the leading order of each term, the equation of motion for $\delta\chi$ from \eqref{EOM_pert} reduces to
\begin{equation}
\underbrace{\epsilon \, \partial_t^2 \delta\chi_1}_{{\cal O}(\epsilon)}
+ \underbrace{\epsilon \, 3 H_0^2 \, \partial_t \delta\chi_1}_{{\cal O}(\epsilon)}
+ \underbrace{\epsilon \beta^2 \left[ V_0'' - \frac{f_0''}{2} \left( \partial_t \phi \right)^2 \right] \delta\chi_1}_{{\cal O}(\epsilon^{-1})}
= \underbrace{\beta f_0' \, \partial_t \phi_0 \, \partial_t \delta\varphi_0}_{{\cal O}(\epsilon^{-1})}
- \underbrace{\frac{\beta f_0 \, \partial_t \phi_0 V_0'}{\Mpl^2 H_0} \, \delta \varphi_0}_{{\cal O}(\epsilon^{-1})} \; ,
\label{EOM_dchi1}
\end{equation}
where again $f_0 \equiv f (\beta \chi_1)$ and $V_0 \equiv V (\beta \chi_1)$.
We observe that the leading order is ${\cal O}(\epsilon^{-1})$, and all the time derivatives of $\delta\chi_1$ drop out. This implies that $\delta\chi_1$ is algebraically given by the other variables without its own dynamics, i.e.~``integrated out,'' and that the above equation plays a role of a constraint rather than E.o.M..
Picking up the ${\cal O}(\epsilon^{-1})$ terms, we find, from \eqref{EOM_dchi1},
\begin{equation}
\delta\chi_1 = \frac{c_{s,0}^{-2} - 1}{\beta \, \partial_t \phi_0}
\left( \frac{f_0}{f_0'} \, \partial_t \delta\varphi_0
- \frac{f_0^2}{f_0'{}^2} \, \frac{V_0'}{\Mpl^2 H_0} \, \delta\varphi_0 \right) \; ,
\label{dchi1}
\end{equation}
where in the last equality $\epsilon$ is absorbed into $\delta\chi_1$, and $c_{s,0}^2$ is defined in \eqref{cs0}. From \eqref{dchi1} it manifests that the leading order of $\delta\chi$ is already at the $\beta^{-1}$ order.
It is worth noting that this EFT reduction would be impossible if $\phi$ is not moving, i.e.~$\partial_t \phi_0 = 0$, or $X_0 = 0$, or if there is no coupling between the two scalar fields, i.e.~$f = {\rm const.}$, as is clear from \eqref{dchi1}.
Now we plug \eqref{dchi1} and its time derivative back into the equation of motion for $\delta\varphi$ in \eqref{EOM_pert} to obtain the EFT equation. The leading order for $\delta\varphi$ is ${\cal O}(\epsilon^{0})$, and, using the background equations, its equation of motion is found as
\begin{equation}
\begin{aligned}
\partial_t^2 \delta\varphi_0 &
+ \left( 3 \, c_{s,0}^2 - \frac{\partial_t c_{s,0}^2}{c_{s,0}^2 H_0} \right) H_0 \, \partial_t \delta\varphi_0
+ \left[ c_{s,0}^2 \, p_0^2
+ \frac{3 \left( 1 + w_0 \right)}{2} \left( 3 \left( 1 + c_{s,0}^2 \right) + \frac{\partial_t c_{s,0}^2}{c_{s,0}^2 H_0} \right)
- \frac{9 \left( 1 + w_0 \right)^2}{2} \right] H_0^2 \, \delta\varphi_0 = 0 \; ,
\end{aligned}
\label{EOM_dphi0}
\end{equation}
where $w_0 \equiv p_0 / \rho_0$ and $p_0 \equiv k / (a H_0)$.
This exactly coincides with the case of the $k$-essence model $P(X)$ with the replacement
\begin{equation}
f_0 \leftrightarrow P_X \; , \qquad
c_{s,0}^2 \leftrightarrow \frac{P_X}{2 X P_X + P_X} \; ,
\label{fcs_corresp}
\end{equation}
which is the correct relation as deduced from \eqref{PX_f} and \eqref{cs0}.
This proves that, starting from the partial UV completion \eqref{action} with the two fields, the single-field $k$-essence EFT is correctly induced as the limit of infinite mass $\beta \to \infty$.
It is a nontrivial reduction, with an arbitrary coupling function $f(\beta\chi)$, on the non-flat, cosmological background and with the gravitational interaction included.

To go to one higher order, integrating out $\delta\chi_2$ and collecting the terms of ${\cal O}(\epsilon)$, the equation of motion for $\delta\varphi_1$ reads
\begin{equation}
\partial_t^2 \delta\varphi_1 + F_0 \, \partial_t \delta\varphi_1 + \Omega^2_0 \, \delta\varphi_1
+ F_1 \, \partial_t \delta\varphi_0 + \Omega^2_1 \, \delta\varphi_0 = 0 \; ,
\label{EOM_dphi1}
\end{equation}
where
\begin{align}
F_0 & = 3 H_0 \, c_{s,0}^2 - \frac{\partial_t c_{s,0}^2}{c_{s,0}^2} \; ,
\label{coeff_F0} \\
\Omega^2_0 & = c_{s,0}^2 \, \frac{k^2}{a^2}
+ \frac{3 \left( 1 + w_0 \right)}{2} \left[ 3 \left( 1 + c_{s,0}^2 \right) + \frac{\partial_t c_{s,0}^2}{c_{s,0}^2 H_0} \right] H_0^2
- \frac{9 \left( 1 + w_0 \right)^2}{2} \, H_0^2 \; ,
\label{coeff_Omega0}\\
F_1 & = \frac{\partial_t \phi_1}{\partial_t \phi_0}
\left[
\frac{3 \left( 1 + w_0 \right)}{2} \, H_0
- \frac{\partial_t c_{s,0}^2}{c_{s,0}^2}
- \frac{\left( \partial_t c_{s,0}^2 \right)^2}{3 \, c_{s,0}^6 H_0}
+ \frac{\partial_t^2 c_{s,0}^2}{3 \, c_{s,0}^4 H_0}
\right] \; ,
\label{coeff_F1} \\
\Omega^2_1 & = \frac{\partial_t \phi_1}{\partial_t \phi_0}
\Bigg[
\frac{9 \left( 1 + w_0 \right) \left( w_0 - c_{s,0}^2 \right)^2}{2 \, c_{s,0}^2} \, H_0^2
- \left( c_{s,0}^2 \, \frac{k^2}{a^2}
- \frac{9 \left( 1 - w_0^2 \right)}{4} \, H_0^2 \right) \frac{\partial_t c_{s,0}}{3 \, c_{s,0}^4 H_0}
+ \frac{1 + w_0}{2 \, c_{s,0}^4} \left( \frac{\left( \partial_t c_{s,0}^2 \right)^2}{c_{s,0}^2 N^2} - \partial_t^2 c_{s,0}^2 \right)
\Bigg] \; .
\label{coeff_Omega1}
\end{align}
As can be seen explicitly from the above, the coefficients $F_0$ and $M_0$ are the same as those for the $0$th order in \eqref{EOM_dphi0}, and $F_1$ and $\Omega^2_1$ are suppressed by $\beta^{-1}$ because of the overall $\partial_t \phi_1$.
Then, combining \eqref{EOM_dphi0} and \eqref{EOM_dphi1}, we obtain the equation of motion for $\delta\varphi$ up to this order,
\begin{equation}
\partial_t^2 \delta\varphi + \left( F_0 + F_1 \right) \partial_t \delta\varphi + \left( \Omega^2_0 + \Omega^2_1 \right) \delta\varphi = 0 \; .
\label{EOMeff_upto1st}
\end{equation}
On the other hand, the effective action for $\delta\varphi$ must have the form, in the Fourier space,
\begin{equation}
S^{(2)}_{\rm eff} = \frac{1}{2} \int \dd t \, \dd^3 k \, a^3 \, T \left(
\left\vert \partial_t \delta\varphi \right\vert^2  - \Omega^2 \left\vert \delta\varphi \right\vert^2
\right) \; ,
\label{action_eff}
\end{equation}
and thus the equation of motion reads
\begin{equation}
\partial_t^2 \delta\varphi
+ \left( 3 H + \frac{\partial_t T}{T} \right) \partial_t \delta\varphi
+ \Omega^2 \, \delta\varphi = 0 \; .
\label{EOMeff_general}
\end{equation}
Comparing the mass terms in \eqref{EOMeff_upto1st} and \eqref{EOMeff_general}, we find
\begin{equation}
\Omega^2 = \Omega^2_0 + \Omega^2_1 \; ,
\label{coeff_Omega}
\end{equation}
and comparing the friction terms in \eqref{EOMeff_upto1st} and \eqref{EOMeff_general} order by order with the use of the background equations, we obtain
\begin{align}
\frac{\partial_t T_0}{T_0} = 3 H_0 \left( c_{s,0}^2 - 1 \right) - \frac{\partial_t c_{s,0}^2}{c_{s,0}^2 N} & \qquad
\implies \qquad
T_0 = \frac{f_0}{c_{s,0}^2} \; ,
\label{coeff_T0} \\
\partial_t \left( \frac{T_1}{T_0} \right)
= \partial_t \left[
\left(
\frac{1 - c_{s,0}^2}{c_{s,0}^2}
+ \frac{\partial_t c_{s,0}^2}{3 \, c_{s,0}^4 H_0}
\right)
\frac{\partial_t \phi_1}{\partial_t \phi_0}
\right] & \qquad
\implies \qquad
T_1 = \frac{f_0}{c_{s,0}^2} \left(
\frac{1 - c_{s,0}^2}{c_{s,0}^2}
+ \frac{\partial_t c_{s,0}^2}{3 \, c_{s,0}^4 N H_0}
\right) \frac{\partial_t \phi_1}{\partial_t \phi_0} \; .
\label{coeff_T1}
\end{align}
Therefore the action up to the $1$st order in $\beta^{-1}$ takes the form \eqref{action_eff} with $T = T_0 + T_1$ and $\Omega^2 = \Omega^2_0 + \Omega^2_1$.
We have now derived the action \eqref{action_eff} and the E.o.M.~\eqref{EOM_dphi1} for $\delta\varphi$, which is the only dynamical degree of freedom in this iterative procedure of EFT reduction.
The above expressions indicate that the current expansion should break down in the limit $c_{s,0}^2 \to 0$ since $T_1 \gg T_0$ and $\Omega_1^2 \gg \Omega_0^2$, i.e.~the EFT description would be invalidated in this limit.
In order to ensure that this result is not an artifact of the choice of the variable, in the following subsection we shall employ another variable that has a more physically transparent meaning in itself.

\subsection{EFT expansion in terms of gauge-invariant energy density perturbation}
\label{subsec:deltaGI}

In deriving the higher-order equations in $\epsilon$, it is instructive to proceed with the gauge-invariant perturbation of the energy density instead of $\delta\varphi$, for a more transparent physical interpretation. We define the energy density by $\rho \equiv n^\mu n^\nu T_{\mu\nu}$, where $n^\mu = ( 1/{\cal N} , \, {\cal N}^i / {\cal N})$ is the normal vector with respect to the $3$-D spatial hypersurface, and ${\cal N}$ and ${\cal N}^i$ are the full-order lapse and shift functions, respectively. Note that ${\cal N} = N = 1$ and ${\cal N}^i = 0$ at the background level, and $\delta {\cal N} = \Phi$ and $\delta {\cal N}^i = \partial_i B / a$ at the linear perturbation with the decomposition given in \eqref{pert_metric}.
Then the linear perturbation of $\rho$ takes the form
\begin{align}
\delta\rho
= f \, \partial_t \phi \, \partial_t \delta\varphi
+ \partial_t \chi \, \partial_t \delta\chi
+ \beta \left[ \frac{f'}{2} \left( \partial_t \phi \right)^2 + V' \right] \delta\chi
- \left[ f \left( \partial_t \phi \right)^2 + \left( \partial_t \chi \right)^2 \right] \Phi \; .
\label{deltarho}
\end{align}
The gauge-invariant combination of $\delta\rho$ we choose to use in this work is the one on a slice comoving with the $\varphi$ direction, defined by
\begin{equation}
\begin{aligned}
\delta\rho_{\rm GI} = \delta\rho
- \frac{\partial_t \bar\rho}{\partial_t \phi} \, \delta\varphi
\; ,
\end{aligned}
\label{deltarho_GI}
\end{equation}
where the background energy density $\bar\rho$ is defined in \eqref{rho_BG}. This choice is natural in the regime of the single-field EFT; on the other hand, once the system recovers to the genuine two-field dynamics, choosing other gauge-invariant quantities may be more appropriate, e.g.~those in \cite{Sasaki:1995aw,Gordon:2000hv}. We stick to the variable \eqref{deltarho_GI} in this work, however, since our primary goal is to demonstrate successful reduction to the EFT and the consistent procedure to compute the corrections to it for large $\beta$.
Using the background equations and \eqref{deltarho} and the Hamiltonian constraint equation in the spatially flat gauge $\Psi = E = 0$, i.e.,
\begin{equation}
\Phi = \frac{f \, \partial_t \phi \, \delta\varphi + \partial_t \chi \, \delta\chi}{2 \Mpl^2 H} \; ,
\label{const_Phi}
\end{equation}
we obtain the expression for the gauge-invariant energy density contrast, given by
\begin{equation}
\begin{aligned}
\delta_{\rm GI}
\equiv \frac{\tilde\rho_{\rm GI}}{\bar\rho}
&
= \frac{1}{\bar\rho} \Bigg(
f \, \partial_t \phi \, \partial_t \delta\varphi
- \frac{f \, \partial_t \phi}{2 \Mpl^2 H} \left[ f \left( \partial_t \phi \right)^2 + \left( \partial_t \chi \right)^2 \right] \delta\varphi
+ \frac{3 H}{\partial_t \phi}
\left[ f \left( \partial_t \phi \right)^2
+ \left( \partial_t \chi \right)^2 \right] \delta\varphi
\\ & \qquad\quad
+ \partial_t \chi \, \partial_t \delta\chi
+ \beta \left[ \frac{f'}{2} \left( \partial_t \phi \right)^2 + V' \right] \delta\chi
- \frac{\partial_t \chi}{2 \Mpl^2 H} \left[ f \left( \partial_t \phi \right)^2 + \left( \partial_t \chi \right)^2 \right] \delta\chi
\Bigg) \; .
\end{aligned}
\label{deltaGI}
\end{equation}
We use this variable as an independent variable in the following analysis, and in fact, since $\delta\chi$ can be integrated out iteratively order by order, it is the only dynamical variable in the EFT expansion.

Let us now perform the expansion in terms of small $\beta^{-1}$. Expanding as in \eqref{back_expansion} for the background and \eqref{pert_expansion} for the perturbations, the gauge-invariant density contrast \eqref{deltaGI} at the leading order reduces to
\begin{align}
\delta_{{\rm GI}, 0} &
= \frac{1}{\rho_0} \,
\frac{2 f_0 X_0}{c_{s,0}^2}
\left(
\frac{\partial_t \delta\varphi_0}{\partial_t \phi_0}
- \frac{f_0}{2 \Mpl^2 H_0} \, \partial_t \phi_0 \, \delta\varphi_0
\right)
+ \frac{3 H_0}{\rho_0} \, f_0 \, \partial_t\phi_0 \, \delta\varphi_0
\label{deltaGI_leading_1} \\ &
= \frac{1 + w_0}{c_{s,0}^2} \left[
\frac{\partial_t \delta\varphi_0}{\partial_t \phi_0}
- \frac{3 \left( 1 + w_0 \right)}{2} \, \frac{H_0}{\partial_t \phi_0} \, \delta\varphi_0
\right]
+ 3 \left( 1 + w_0 \right) \, \frac{H_0}{\partial_t \phi_0} \, \delta\varphi_0
\; ,
\label{deltaGI_leading_2}
\end{align}
after using the constraint equations \eqref{EOMback_chi_0} for $\chi_1$ and \eqref{dchi1} for $\delta\chi_1$ and replacing $f_0$ in favor of $w_0$ in the second equality. Note that this expression is exactly the same as the equivalent variable in the single-field $k$-essence model.
Following the procedure summarized in Appendix \ref{app:action}, we obtain the quadratic action for $\delta_{{\rm GI},0}$ in the Fourier space, given by
\begin{equation}
\begin{aligned}
S^{(2)}_0 = \frac{1}{2} \int \dd t \, \dd^3k \, a^3
\tilde{T}_0
\left(
\left\vert \partial_t \delta_{{\rm GI},0} \right\vert^2
- \tilde{\Omega}_0^2 \left\vert \delta_{{\rm GI},0} \right\vert^2
\right) \; ,
\end{aligned}
\label{action_eff_deltaGI}
\end{equation}
where again $p_0 = k/(aH_0)$, and
\begin{align}
    \tilde{T}_0 &= \frac{3 \Mpl^2}{\left( 1 + w_0 \right) p_0^2} \; , \qquad
    \tilde{\Omega}^2_0 = \frac{3 \Mpl^2}{\left( 1 + w_0 \right) p_0^2}
    \left(
    c_{s,0}^2 \, p_0^2 + 15 + 9 \, c_{s,0}^2 - 21 \left( 1 + w_0 \right)
    + \frac{9 \left( 1 + w_0 \right)^2}{2}
    \right) H_0^2 \; .
    \label{coeff_tildeOmega0}
\end{align}
One can easily confirm that this expression can be exactly recovered by starting from the corresponding $k$-essence theory, with the correct identification \eqref{fcs_corresp}. This concludes the successful reduction from the two-field theory to the single-field EFT as the leading order in the expansion $\beta \to \infty$.

Proceeding to the first-order in the expansion of small $\beta^{-1}$, the effective action is given by \eqref{action_eff} together with the coefficients \eqref{coeff_Omega0}, \eqref{coeff_Omega1}, \eqref{coeff_T0} and \eqref{coeff_T1}.
In this computation, we use the same variable as in \eqref{deltaGI_leading_2} (with the replacement $\delta_{{\rm GI},0} \to \delta_{{\rm GI}}$), only aiming for the calculations of the corrections to the EFT dynamics from the higher order, instead of making observable predictions.
Including up to the first sub-leading order in $\epsilon = {\cal O}(\beta^{-1})$, we find the action of the form
\begin{equation}
S^{(2)}_{0 \& 1} = \frac{1}{2} \int \dd t \, \dd^3 k \, a^3 \left( \tilde{T}_0 + \tilde{T}_1 \right) \left[
\left\vert \partial_t \delta_{\rm GI} \right\vert^2
- \left( \tilde\Omega^2_0 + \tilde\Omega^2_1 \right) \left\vert \delta_{\rm GI} \right\vert^2
\right] \; ,
\label{action_eff_deltaGI_upto1}
\end{equation}
where $\tilde{T}_0$ and $\tilde{\Omega}_0^2$ are the same as given above, and
\begin{align}
    \tilde{T}_1 & = \frac{3 \Mpl^2}{\left( 1 + w_0 \right) p_0^4} \, \frac{\partial_t \phi_1}{\partial_t \phi_0}
    \Bigg[ - p_0^2 + \frac{1}{c_{2,0}^2} \left( p_0^2 - 9 \left( 1 + w_0 \right) + \frac{27 \left( 1 + w_0 \right)^2}{4} \right)
    - \frac{9 \left( 1 + w_0 \right) w_0^2}{2 \, c_{s,0}^4}
    \nonumber\\
    & \qquad\qquad\qquad\qquad\quad
    - 3 \left( 1 - \frac{2 \, p_0^2}{9 \, c_{s,0}^2} - \frac{1 + w_0}{2 \, c_{s,0}^2} + \frac{1 - w_0^2}{4 \, c_{s,0}^4} \right) \frac{\partial_t c_{s,0}^2}{c_{s,0}^2 H_0}
    - \frac{\left( \partial_t c_{s,0}^2 \right)^2}{c_{s,0}^6 H_0^2}
    + \frac{\partial_t^2 c_{s,0}^2}{c_{s,0}^4 H_0^2}
    \Bigg] \; .
\end{align}
The full expression of $\tilde{\Omega}_1^2$ is rather lengthy and is not important for our purpose, and thus we only write the expression with constant $c_{s,0}^2$ here, giving
\begin{align}
\tilde\Omega^2_1 & = \frac{3 H_0^2 \left( 1 + w_0 \right)}{16 \, p^2} \, \frac{\partial_t \phi_1}{\partial_t \phi_0}
\Bigg[ - 108 \left( 1 - w_0 \right) \left( 1 + 3 \, w_0 \right)
- 8 \left( 1 + 9 \, w_0 \right) p^2
\nonumber\\ & \qquad
+ \frac{45 \left( 1 + 3 \, w_0 \right)^2 \left( 1 - w_0 \right)
+ 8 \left( 5 - 7 \left( 1 + w_0 \right) + 3 \left( 1 + w_0 \right)^2 \right) p^2}{c_{s,0}^2}
- \frac{18 \, w_0^2 \left( 1 - w_0 \right) \left( 7 + 9 \, w_0 \right)}{c_{s,0}^4}
\Bigg]  \; .
\end{align}
From this expression, it is evident that the EFT expansion breaks down for $c_{s,0}^2 \to 0$, as it drives $\tilde{T}_1 \gg \tilde{T}_0$ and $\tilde{\Omega}^2_1 \gg \tilde{\Omega}^2_0$, which is expected by a general argument of EFT \cite{Cheung:2007st}.

In this section, therefore, we have explicitly shown that the EFT reduction is successfully done as the leading order in the limit $\beta \to \infty$, given in the $0$th-order \eqref{action_eff} for $\delta\varphi$ and \eqref{action_eff_deltaGI} for the gauge-invariant density contrast $\delta_{\rm GI}$, and that the sub-leading corrections can be unambiguously derived by iteratively expanding the orders of small $\beta^{-1}$, found in the $1$st-order \eqref{action_eff} for $\delta\varphi$ and \eqref{action_eff_deltaGI_upto1} for $\delta_{\rm GI}$.
In passing, we also observe that the $c_{s,0}^2 \to 0$ limit triggers the departure from the EFT description.

\section{Summary and discussion}
\label{sec:discussion}

The class of $k$-essence models are widely used in the context of cosmological applications, for both early- and late-time accelerated expansion. The dynamics of the scalar field(s) in these models drive the expansion and lead to the predictions of inflationary observables as well as the fate of the universe.
While the $k$-essence has attracted much attention in this respect, it has been pointed out that models of its shift-symmetric version generically form caustic singularities in the spacetime regions where a planar-symmetric configuration is well respected \cite{Babichev:2016hys,Mukohyama:2016ipl,deRham:2016ged}. Two classes of shift-symmetric $k$-essence are known to be free from the caustics, namely the standard canonical scalar \cite{Babichev:2016hys} and the scalar field with the DBI-type kinetic term \cite{Mukohyama:2016ipl}.
In this paper, with this knowledge in mind, we have studied two-field completions of some general classes of shift-symmetric single-field $k$-essence models for those two cases. To this end, we have introduced a parameter $\beta$ that controls the mass scale of the second field $\chi$, so that the single-field EFT description should be recovered in the limit $\beta \to \infty$, equivalently $m_\chi \to \infty$, by integrating out the second field.

In Sec.~\ref{sec:twofield}, we have introduced the class of $k$-essence we consider as an EFT and then its (partially) UV-completed model by promoting a second field to a dynamical degree of freedom on a curved field space. We have exemplified the flat, hyperboloidal and spheroidal geometry of the field space. The completion has been done both for the linear kinetic terms and for the DBI-type kinetic terms, and in each case, we have shown that the two-field model is formally reduced to the expected single-field $k$-essence EFT in the $\beta \to \infty$ limit.

Sec.~\ref{sec:planar} has been devoted to the explicit demonstration of the caustic formation in the single-field EFT and of its resolution by the two-field hyperboloidal field space, by performing numerical integrations. To our knowledge, this is the first numerical illustration of the formation and resolution of caustics in a $k$-essence model and its UV-completed theory.
From the numerical result, it is evident that the dynamics of the EFT evolves into the formation of caustics as the second derivative of the scalar field, $\partial^2\varphi$, diverges. This singularity is resolved in the two-field case by transferring the energy to the second field $\chi$ prior to the caustic formation, and consequently the second derivative $\partial^2\varphi$ is smoothed out. This is the moment when the EFT description breaks down and the system turns into a full two-field dynamics.
For a smaller value of the controlling parameter $\beta$, the system deviates from the EFT at a lower energy scale, i.e.~at an earlier time during the evolution. This expectation has indeed been confirmed in the numerical calculation, and consequently the shape of the wave stays smoother for a smaller $\beta$ than for a larger one.
This completes the demonstration of the partial UV completion of the shift-symmetric $k$-essence, with the use of a curved field space in the UV sector.

In Sec.~\ref{sec:cosmology}, we have then considered the above-verified UV model in view of cosmological applications. We have first derived the background equations on the flat FLRW metric. Expanding for small $\beta^{-1}$ and collecting the leading-order terms in each equation, we have observed that the leading order of the heavy field $\chi$ is in fact ${\cal O}(\beta^{-1})$ to derive the ${\cal O}(\beta^0)$ equations for the light field $\varphi$. The resulting leading EFT equations have been shown to exactly reproduce those obtained starting from the corresponding $k$-essence model $P(X)$. The sub-leading corrections can also be deduced iteratively in a straightforward manner in the small $\beta^{-1}$ expansion.
Turning to the cosmological perturbations around the background, we have conducted a detailed study of the scalar sector, as the vector and tensor perturbations are unchanged from the standard canonical single-field model. As in the background calculation, we have first derived the equations of the genuine two-field system and then expanded them for small $\beta^{-1}$. In this expansion, $\delta\chi$ can be iteratively integrated out order by order, and the system is effectively reduced to a single-field one at each order. This master equation of the linear perturbation indeed reproduces the corresponding $k$-essence equation as the leading order in $\beta^{-1} \to 0$. The higher-order corrections have again been computed iteratively without ambiguity. For a transparent physical interpretation, we have converted the single variable to the gauge-invariant density contrast $\delta_{\rm GI} = \delta\rho_{\rm GI} / \bar\rho$ and obtained the quadratic action in terms of $\delta_{\rm GI}$ using the procedure summarized in Appendix \ref{app:action}. Looking at the leading and first-order contributions to the action, we have observed that this expansion breaks down in the limit of vanishing sound speed $c_s^2 \to 0$, which is consistent with the discussion in the language of EFT seen in e.g.~\cite{Cheung:2007st}.
Therefore, in Sec.~\ref{sec:cosmology}, we have provided the explicit demonstration that the correct reduction from the two-field model to the single-field EFT as the $\beta^{-1} \to 0$ limit, with the gravity taken into account, that the sub-leading terms can be iteratively computed, and that the cutoff scale of the EFT description decreases arbitrarily in the limit $c_s^2 \to 0$.

Our detailed analysis is focused primarily on the completion by the linear kinetic terms (with a curved field space). It can be extended to the case of the DBI-type kinetic terms in a straightforward, but perhaps more tedious, manner. We expect the main qualitative conclusions in Secs.~\ref{sec:planar} and \ref{sec:cosmology} to be unchanged.
Also, as shown in \cite{Mukohyama:2016ipl}, the avoidance of caustics in a planar-symmetric configuration only requires an appropriate choice of the $k$-essence part in the Horndeski theory \cite{Horndeski:1974wa,Deffayet:2011gz,Kobayashi:2011nu}. Thus the UV completion introduced in Sec.~\ref{sec:twofield} of this work should be applicable in the presence of the higher-order (shift-symmetric) Horndeski terms. Extending the computation done in Sec.~\ref{sec:cosmology} to such Horndeski models is also of interest for further investigation.
Finally, our computation in Sec.~\ref{sec:cosmology} concentrates on the EFT reduction from the UV theory. It would be exciting to see how the $\beta^{-1}$ suppressed contributions, i.e.~the effects from the UV, modify the observables such as inflationary predictions that are computed only from the single-field EFT.
We leave these considerations to upcoming studies and would like to come back to these issues in the near future.

\section*{acknowledgement}
R.N.~is grateful to Elisa G.M.~Ferreira and Motoo Suzuki for casual discussions on the topic.
The work of S.M. was supported in part by Japan Society for the Promotion of Science Grants-in-Aid for Scientific Research No.~17H02890, No.~17H06359, and by World Premier International Research Center Initiative, MEXT, Japan.

\appendixpage
\appendix

\section{Change of variables with derivatives and derivation of its action}
\label{app:action}

In this appendix, we formulate the derivation of quadratic action/Lagrangian in terms of the variable that consists of a linear combination of the original variable and its first time derivative. This technique is introduced in Appendix B of \cite{DeFelice:2015moy} (see also \cite{Gumrukcuoglu:2016jbh}), and here we keep track of the time dependence of all the coefficients.
For our purpose, i.e.~derivation in the Fourier space and on an isotropic and homogeneous background, it suffices to consider a one-variable classical-mechanical system of a quadratic Lagrangian
\begin{equation}
L = \frac{T}{2} \, \dot{q}^2 - \frac{M}{2} \, q^2 \; ,
\label{lag_original}
\end{equation}
where $q$ is the physical variable, dot denotes derivative with respect to time, and $T$ and $M$ are in general functions of time.
We aim to describe the dynamics using another variable, say
\begin{equation}
Q = C \, \dot q + D \, q \; ,
\end{equation}
instead of $q$. To this end, we rewrite the Lagrangian \eqref{lag_original} as
\begin{equation}
\begin{aligned}
L & = \frac{T}{2 C^2} \left( C \, \dot q + D \, q \right)^2
- \frac{1}{2} \left[ M + \frac{T D^2}{C^2} - \partial_t \left( \frac{T D}{C} \right) \right] q^2
- \partial_t \left( \frac{T D}{2 \, C} \, q^2 \right)
\\ &
= \frac{T}{2 C^2} \left[ 2 Q \left( C \, \dot q + D \, q \right) - Q^2 \right]
- \frac{1}{2} \left[ M + \frac{T D^2}{C^2} - \partial_t \left( \frac{T D}{C} \right) \right] q^2
- \partial_t \left( \frac{T D}{2 \, C} \, q^2 \right) \; .
\end{aligned}
\end{equation}
Varying this with respect to $Q$ and plugging the expression for $Q$ back into the Lagrangian, it is clear to that the original action \eqref{lag_original} is restored up to total derivatives.
Now, we further manipulate the above expression as, by completing the square for $q$,
\begin{equation}
\begin{aligned}
L &
= - \frac{1}{2} \left[ M + \frac{T D^2}{C^2} - \partial_t \left( \frac{T D}{C} \right) \right]
\left[ q - \frac{\frac{T D}{C^2} \, Q
- \partial_t \left( \frac{T}{C} \, Q \right)}{M + \frac{T D^2}{C^2} - \partial_t \left( \frac{T D}{C} \right)}  \right]^2
\\ & \quad
+ \frac{1}{2} \, \frac{\left[ \frac{T D}{C^2} \, Q
- \partial_t \left( \frac{T}{C} \, Q \right) \right]^2}{M + \frac{T D^2}{C^2} - \partial_t \left( \frac{T D}{C} \right)}
- \frac{T}{2 C^2} \, Q^2
+ \partial_t \left( \frac{T}{C} \, Q q - \frac{T D}{2 \, C} \, q^2 \right)
\; .
\end{aligned}
\label{lag_2}
\end{equation}
Provided
\begin{equation}
M + \frac{T D^2}{C^2} - \partial_t \left( \frac{T D}{C} \right) \ne 0 \; ,
\end{equation}
we can vary the action with respect to $q$ and solve an algebraic equation for $q$. Then the first line of \eqref{lag_2} vanishes, and the Lagrangian becomes
\begin{equation}
\begin{aligned}
L & = \frac{1}{2} \, \frac{\left[ \frac{T}{C} \, \dot{Q} - \frac{T D}{C^2} \, Q
+ \partial_t \left( \frac{T}{C} \right) Q \right]^2}{M + \frac{T D^2}{C^2} - \partial_t \left( \frac{T D}{C} \right)}
- \frac{T}{2 C^2} \, Q^2
+ \left( \mbox{total derivatives} \right)
\\ &
= \frac{1}{2} \, \frac{T^2 \, \dot{Q}^2
}{M C^2 + T D^2 - C^2 \partial_t \left( \frac{T D}{C} \right)}
\\ & \quad
- \frac{1}{2} \left\{
\frac{T M
+ T D \, \partial_t \left( \frac{T}{C} \right)
- \frac{T^2}{C} \, \dot{D}
- C^2 \left[ \partial_t \left( \frac{T}{C} \right) \right]^2
}{M C^2 + T D^2 - C^2 \partial_t \left( \frac{T D}{C} \right)}
- \partial_t \left[
\frac{\frac{T^2 D}{C} - T C \, \partial_t \left( \frac{T}{C} \right)}{M C^2 + T D^2 - C^2 \partial_t \left( \frac{T D}{C} \right)}
\right]
\right\} Q^2
+ \left( \mbox{total derivatives} \right)'
\; .
\end{aligned}
\label{lag_trans}
\end{equation}
where prime is just a bookmark to note that it is a total derivative different from the previous line.
This Lagrangian is fully expressed in terms of $Q$, while its physical content is completely equivalent to the original action \eqref{lag_original}, at least classically.

\subsection{Relation to Canonical Transformation}

In this subsection, we show that the above transformation of the Lagrangian is indeed a canonical transformation, as a consistency check.
From \eqref{lag_original}, the conjugate momentum of $q$ is
\begin{equation}
p \equiv \frac{\delta L}{\delta \dot{q}} = T \dot{q}
\; ,
\end{equation}
and the Hamiltonian is
\begin{equation}
H = p \dot{q} - L = \frac{p^2}{2T} + \frac{M}{2} \, q^2 \; .
\label{Hamil_original}
\end{equation}
The Poisson bracket with respect to the $\{ q , \, p \}$ canonical pair is defined by
\begin{equation}
\{ X , \, Y \} \equiv \frac{\delta X}{\delta q} \, \frac{\delta Y}{\delta p} - \frac{\delta X}{\delta p} \, \frac{\delta Y}{\delta q} \; ,
\label{Poisson}
\end{equation}
and it is obvious that $\{ q , p \} = 1$.

On the other hand, from \eqref{lag_trans}, the conjugate momentum of $Q$ is
\begin{equation}
P \equiv \frac{\delta L}{\delta \dot Q} = \frac{T^2 \, \dot{Q}
}{M C^2 + T D^2 - C^2 \partial_t \left( \frac{T D}{C} \right)}
\; ,
\end{equation}
and the Hamiltonian is
\begin{equation}
\begin{aligned}
H & = P \dot Q - L
\\ &
= \frac{M C^2 + T D^2 - C^2 \partial_t \left( \frac{T D}{C} \right)}{2 T^2} \, P^2
\\ & \quad
+ \frac{1}{2} \left\{
\frac{T M
+ T D \, \partial_t \left( \frac{T}{C} \right)
- \frac{T^2}{C} \, \dot{D}
- C^2 \left[ \partial_t \left( \frac{T}{C} \right) \right]^2
}{M C^2 + T D^2 - C^2 \partial_t \left( \frac{T D}{C} \right)}
- \partial_t \left[
\frac{\frac{T^2 D}{C} - T C \, \partial_t \left( \frac{T}{C} \right)}{M C^2 + T D^2 - C^2 \partial_t \left( \frac{T D}{C} \right)}
\right]
\right\} Q^2 \; .
\end{aligned}
\label{Hamil_trans}
\end{equation}
From the construction, the transformation $( q , p ) \to( Q , P )$ is a canonical one, and we show this explicitly below.

From the original Hamiltonian \eqref{Hamil_original}, the Euler-Lagrange equations are
\begin{equation}
\dot{q} = \left\{ q , H \right\} = \frac{p}{T} \; , \qquad
\dot{p} = \left\{ p , H \right\} = - M q \; .
\end{equation}
Using this, $Q$ and $P$ can be expressed in terms of $q$ and $p$ as
\begin{equation}
\begin{aligned}
Q & = \frac{C}{T} \, p + D \, q \; , \qquad
P
= \frac{T}{M C^2 + T D^2 - C^2 \partial_t \left( \frac{T D}{C} \right)}
\left[
\left( D + T \, \partial_t \left( \frac{C}{T} \right) \right) p
+ \left( T \dot{D} - MC \right) \, q
\right] \; .
\end{aligned}
\end{equation}
Then the Poisson bracket \eqref{Poisson} of $Q$ and $P$ reads
\begin{equation}
\begin{aligned}
\left\{ Q , \, P \right\}
= D \, \frac{T \left( D + T \partial_t \left( \frac{C}{T} \right) \right)}{M C^2 + T D^2 - C^2 \partial_t \left( \frac{T D}{C} \right)}
- \frac{C}{T} \, \frac{T \left( T \dot{D} - MC \right)}{M C^2 + T D^2 - C^2 \partial_t \left( \frac{T D}{C} \right)}
= 1 \; .
\end{aligned}
\end{equation}
Therefore, $( q , p) \to ( Q , P )$ is a canonical transformation, and we can treat $( Q , P )$ as a canonical pair to compute the Poisson bracket,
\begin{equation}
\{ X , \, Y \}' \equiv \frac{\delta X}{\delta Q} \, \frac{\delta Y}{\delta P} - \frac{\delta X}{\delta P} \, \frac{\delta Y}{\delta Q} \; ,
\end{equation}
because
\begin{equation}
\begin{aligned}
\left\{ X , \, Y \right\} & =
\frac{\delta X}{\delta q} \, \frac{\delta Y}{\delta p} - \frac{\delta X}{\delta p} \, \frac{\delta Y}{\delta q} \\
& = \left( \frac{\delta X}{\delta Q} \, \frac{\delta Q}{\delta q} + \frac{\delta X}{\delta P} \, \frac{\delta P}{\delta q} \right)
\left( \frac{\delta Y}{\delta Q} \, \frac{\delta Q}{\delta p} + \frac{\delta Y}{\delta P} \, \frac{\delta P}{\delta p} \right)
- \left( \frac{\delta X}{\delta Q} \, \frac{\delta Q}{\delta p} + \frac{\delta X}{\delta P} \, \frac{\delta P}{\delta p} \right)
\left( \frac{\delta Y}{\delta Q} \, \frac{\delta Q}{\delta q} + \frac{\delta Y}{\delta P} \, \frac{\delta P}{\delta q} \right)
\\ &
= \frac{\delta X}{\delta Q} \, \frac{\delta Y}{\delta P} - \frac{\delta X}{\delta P} \, \frac{\delta Y}{\delta Q}
= \left\{ X , \, Y \right\}' \; .
\end{aligned}
\end{equation}
Also it is then immediate to see, by taking time derivative of $Q$ and $P$ using the Poisson brackets $\{ \bullet , \bullet\}$ with respect to $(q,p)$,
\begin{equation}
\begin{aligned}
\dot Q & = \partial_t \left( \frac{C}{T} \right) p + \dot{D} \, q
+ \frac{C}{T} \left\{ p , \, H \right\} + D \left\{ q , \, H \right\}
\\ &
= \frac{M C^2 + T D^2 - C^2 \partial_t \left( \frac{T D}{C} \right)}{T^2} \, P \; ,
\\
\dot P & =
\partial_t \left[ \frac{T \left( D + T \, \partial_t \left( \frac{C}{T} \right) \right)}{M C^2 + T D^2 - C^2 \partial_t \left( \frac{T D}{C} \right)} \right] p
+ \partial_t \left[ \frac{T \left( T \dot{D} - MC \right)}{M C^2 + T D^2 - C^2 \partial_t \left( \frac{T D}{C} \right)} \right] q
\\ & \quad
+ \frac{T \left( T D + T \, \partial_t \left( \frac{C}{T} \right) \right)}{M C^2 + T D^2 - C^2 \partial_t \left( \frac{T D}{C} \right)} \left\{ p , \, H \right\}
+ \frac{T \left( T \dot{D} - MC \right)}{M C^2 + T D^2 - C^2 \partial_t \left( \frac{T D}{C} \right)} \left\{ q , \, H \right\}
\\ &
= - \left\{ \frac{T M + T D \, \partial_t \left( \frac{T}{C} \right) - \frac{T^2}{C} \, \dot{D}
- C^2 \, \left[ \partial_t \left( \frac{T}{C} \right) \right]^2
}{M C^2 + T D^2 - C^2 \partial_t \left( \frac{T D}{C} \right)}
- \partial_t \left[ \frac{\frac{T^2}{C} \left( D + T \, \partial_t \left( \frac{C}{T} \right) \right)}{M C^2 + T D^2 - C^2 \partial_t \left( \frac{T D}{C} \right)} \right] \right\} Q \; .
\end{aligned}
\end{equation}
These equations are precisely the Euler-Lagrange equations that can be obtained from the transformed Hamiltonian \eqref{Hamil_trans}.
Therefore, the dynamics of the $( q ,p )$ system is reproduced by that of the $( Q , P )$ system in the exact manner.
This concludes the equivalence of the two systems.

\bibliography{caustics_complete}

\end{document}